\documentclass{article}
\usepackage[margin=1in]{geometry}
\usepackage{amsmath}
\usepackage{amssymb}
\usepackage{graphicx}
\usepackage{booktabs}
\usepackage{todonotes}
\usepackage[authoryear]{natbib}
\usepackage{longtable}
\usepackage[table]{xcolor}

\usepackage{subcaption}

\usepackage[most]{tcolorbox}
\newtcolorbox{myquotebox}{%
  colback=blue!5,
  colframe=blue!5, %blue!60!black,
  boxrule=0.8pt,
  arc=4pt,
  outer arc=4pt,
  boxsep=6pt,
  left=8pt,
  right=8pt,
  top=6pt,
  bottom=6pt,
  fontupper=\sffamily
}
\newcommand{\attributed}[1]{\hspace{1em}--- #1}

\title{Agentic AI in Engineering and Manufacturing: Industry Perspectives on Utility, Adoption, Challenges, and Opportunities}

\author{Kristen M. Edwards, Maxwell Bauer, Claire Jacquillat, A. John Hart, Faez Ahmed}
\date{March 2026}

\begin{document}

% \makeatletter
% \renewcommand\@fnsymbol[1]{}
% \makeatother

\maketitle
\begingroup
\renewcommand{\thefootnote}{}
\footnotetext{This work was funded and supported by the MIT Initiative for New Manufacturing.}
\endgroup

\vspace{-1cm}

\section*{Abstract}
This work examines how AI, especially agentic systems, is being adopted in engineering and manufacturing workflows, what value it provides today, and what is needed for broader deployment. This is an exploratory and qualitative state-of-practice study grounded in over 30 interviews across four stakeholder groups (large enterprises, small/medium firms, AI developers, and CAD/CAM/CAE vendors). We find that near-term AI gains cluster around structured, repetitive work and data-intensive synthesis, while higher-value agentic gains come from orchestrating multi-step workflows across tools. Adoption is constrained less by model capability than by fragmented and machine-unfriendly data, stringent security and regulatory requirements, and limited API-accessible legacy toolchains. Reliability, verification, and auditability are central requirements for adoption, driving human-in-the-loop frameworks and governance aligned with existing engineering reviews. 
Beyond technical barriers there are also organizational ones: a persistent AI literacy gap, cultural heterogeneity, and governance structures that have not yet caught up with agentic capabilities. Together, the findings point to a staged progression of AI utility from low-consequence assistance toward higher-order automation, as trust, infrastructure, and verification mature. This highlights key breakthroughs needed, including integration with traditional engineering tools and data types, robust verification frameworks, and improved spatial and physical reasoning.

% \newpage
% \tableofcontents
% \newpage
\begin{figure}[h]
    \centering
    \includegraphics[width=\linewidth]{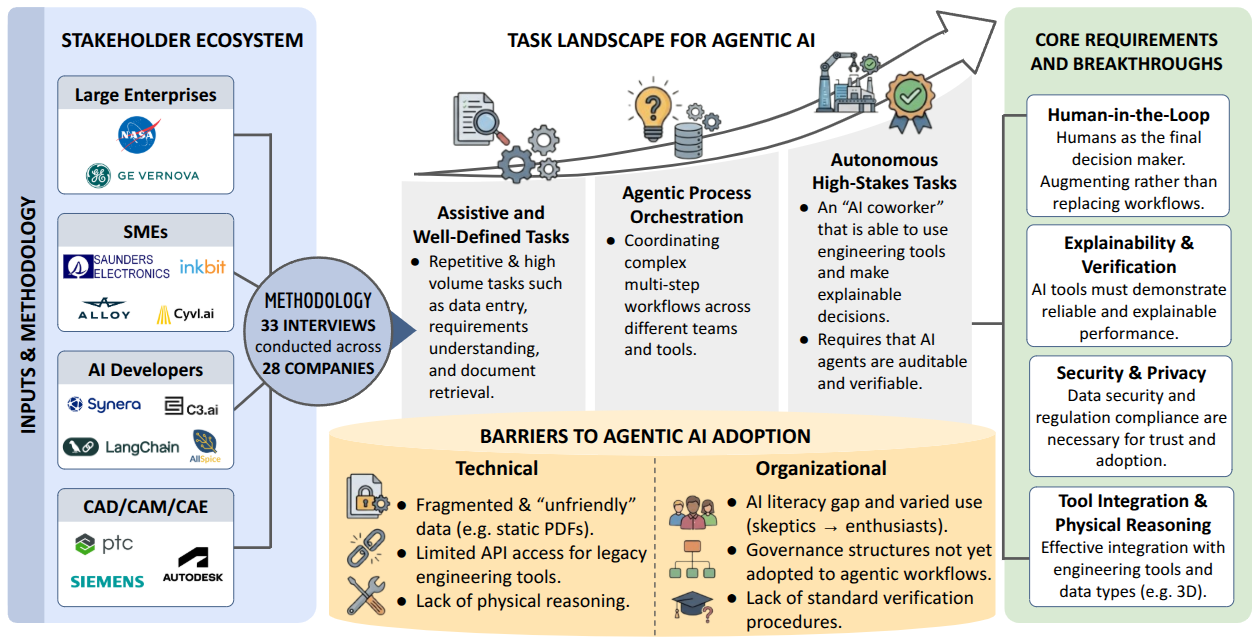}
    \caption{Overall findings for agentic AI adoption in engineering and manufacturing.}
    \label{fig:overall}
\end{figure}

\section{Introduction}

% \todo[inline]{Add an appendix on interview guide. How were participants recruited? referaals or puposive sampling? What was the interview protocol? mention details like interview duration, recording/transcription. add a concise Analysis subsection: how notes/transcripts were handled, how themes were developed, and how the team assessed saturation. How were disagreements resolved, and how did the team avoid confirmation bias?}

%\todo[inline]{Faez: How do you distinguish agentic AI from workflow automation + LLM in your analysis, and what autonomy level are your conclusions really about? Providing a definition of agentic AI (inspired by the ones provided by Anthropic/OpenAI) and then discussing different flavors may help. This could be a discussion or an initial clarification.}

% \todo[inline]{The introduction is long and may dilute the core contribution. 3 pages about manufacturing before getting to the key messages of the paper is too much.}

% Say that this is a state of practice paper or an exploratory qualitative study

\subsection{The Engineering and Manufacturing Landscape}

Engineering and manufacturing are foundational to the global economy and national competitiveness. In 2022, global manufacturing added \$15.0 trillion of value in constant 2015 dollars (17.5\% of global GDP) according to the United Nations Statistics Division ~\citep{NIST2024AMS60016}. The top two contributing countries, China and the United States, accounted together for approximately 45\% of that output. In the United States specifically, engineering occupations are projected to grow ``faster than average for all occupations from 2024 to 2034''~\citep{bls2023}. The importance of the manufacturing sector is clear: the value-added output of US manufacturing was approximately \$2.90 trillion (annualized) in Q1 2025, accounting for 9.7\% of GDP as wel as 35\% of productivity growth, and every \$1.00 spent on manufacturing adds about \$2.64 to the overall U.S. economy~\citep{NAM2025FactsAboutManufacturing, Bistarkey2022ManufacturingEcosystem}. %Furthermore, each manufacturing job supports approximately 4.8 additional jobs through indirect and induced effects ~\citep{NAM2025FactsAboutManufacturing}. 

\paragraph{Resurgence in U.S. Manufacturing}

\begin{figure}[h]
    \centering
    \includegraphics[width=0.8\linewidth]{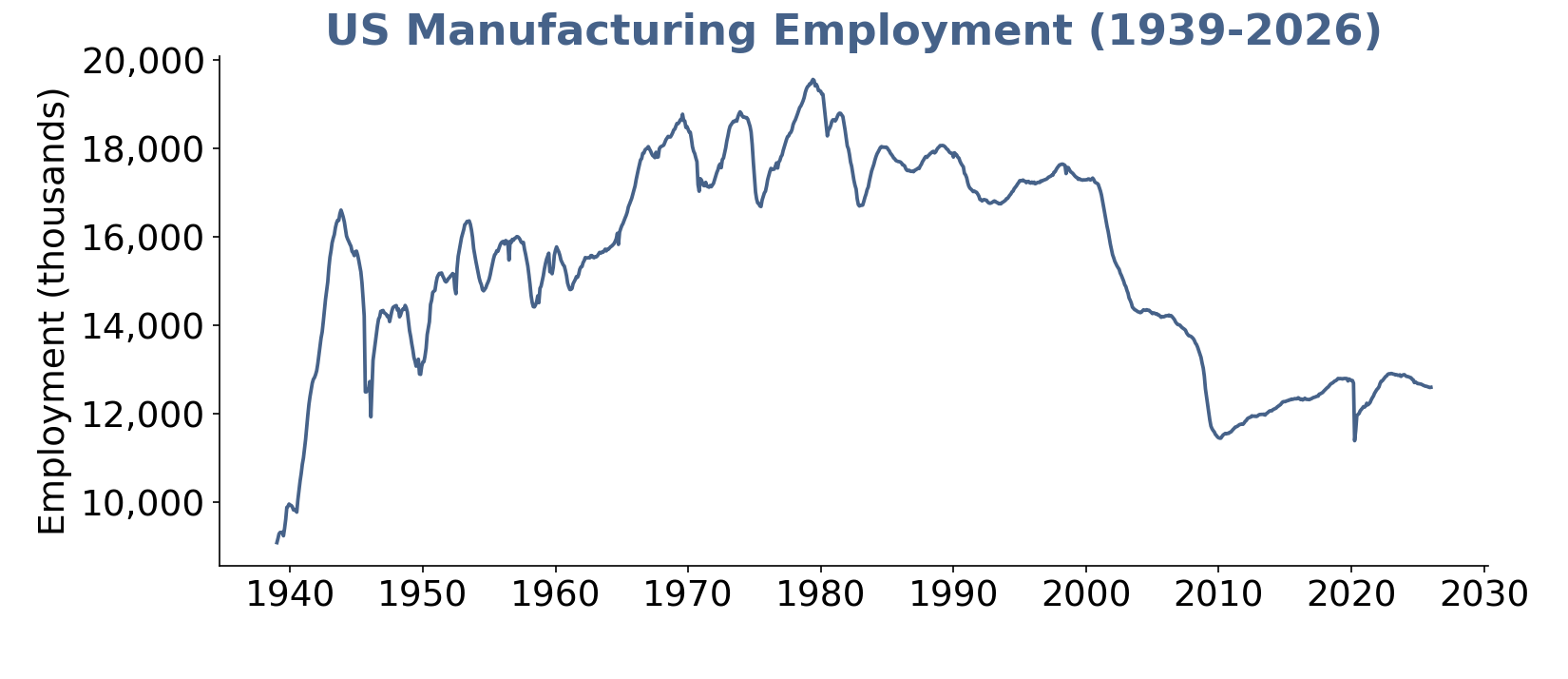}
    \caption{U.S. manufacturing employment in thousands of people. Source of data: U.S. Bureau of Labor Statistics.}
    \label{fig:bls-mfg-employment}
\end{figure}

U.S. manufacturing employment peaked at 19.6 million in 1979 before falling to under 12 million around 2010 (Figure~\ref{fig:bls-mfg-employment}). Since then, employment has grown steadily, reaching just under 13 million in 2024 ~\citep{bls2020, bahr2025, deloitte2024}. Concurrently, U.S. manufactured goods exports hit a record \$1.65 trillion in 2024, and dollars invested in new manufacturing facilities nearly tripled between 2020 and 2024 ~\citep{NAM2025FactsAboutManufacturing, deloitte2024}. Legislation like the Infrastructure Investment and Jobs Act (IIJA), the Inflation Reduction Act (IRA), and the Creating Helpful Incentives to Produce Semiconductors (CHIPS) Act represent over \$430 billion invested and include announcements of more than 234,000 new manufacturing jobs to be created~\citep{deloitte2024}. 

Although there has been steady growth in the workforce since 2010~\citep{bls2020}, a 2024 Deloitte report suggests that the U.S. may face a shortage of 1.9 million manufacturing engineers by 2033~\citep{deloitte2024}. Meeting the rising manufacturing demand has become an increasing concern, underscoring the need for innovation and efficiency.

\paragraph{Reshoring and Nearshoring Trends}

One of the reasons for the growing need for a domestic manufacturing workforce is that the manufacturing operating model is shifting from long, optimized-for-cost supply chains to more regional, resilient, and digitally enabled networks. COVID-19, tariffs, and geopolitical instability have been driving factors for this shift~\citep{xometry2026, Yellen2023TreasuryRemarks}. Across hundreds of leaders surveyed by the World Economic Forum and Kearney, more than 90\% report prioritizing regionalization, and nearly two-thirds plan to qualify a majority of direct spend from a second source in a second region~\citep{WEF2024DisruptionToOpportunity}. Furthermore, Xometry and Thomas's \textit{2026 Manufacturing Outlook} reports that 45\% of global executives are actively working to reshore
facilities and 29\% have already done so~\citep{xometry2026}.

According to Forbes, over \$400~billion in advanced manufacturing projects in North America have been pledged through 2030~\citep{Forbes2024ResurgenceUSManufacturing}. Automotive manufacturing leads this shift, with companies such as BMW and Tesla expanding North American production to improve resilience, agility, and quality control. Enabled by the U.S.–Mexico–Canada Agreement (USMCA), firms are increasingly locating facilities in Mexico and the southern United States to reduce supply chain risks and shorten lead times. This movement is reinforced by automation and AI adoption, which help offset higher labor costs while creating new opportunities in advanced manufacturing, logistics, and industrial real estate~\citep{Forbes2024ResurgenceUSManufacturing}.

This represents a structural pivot intended to absorb volatility while protecting service levels and growth. This evolution extends traditional “cost-only” location logic: country readiness now depends on capabilities in resilience, digitalization, sustainability, talent, and ecosystem orchestration, not just labor costs~\citep{WEF2024BeyondCost}.

\paragraph{Workforce Needs}
Despite strong growth in global manufacturing, multiple reports warn that the industrial workforce may be insufficient to meet rising demand. Persistent attrition and skills shortages affect frontline and technical roles, while employment has declined even as output has risen~\citep{WEF2025EmpoweringFrontlines,NIST2024AMS60016}. Technology is advancing faster than the workforce can adapt, with skill obsolescence posing a further constraint on productivity~\citep{ieee2024}. Together, these trends underscore the urgency of improving efficiency through digital transformation, automation, and AI-enabled systems to sustain manufacturing growth with a limited labor base. Parallel analyses in the World Economic Forum's \textit{Beyond Cost} and \textit{From Disruption to Opportunity} white papers underscore the same challenge at a systemic scale, linking national manufacturing competitiveness to labor shortages, skills erosion, and the aging workforce~\citep{WEF2024BeyondCost,WEF2024DisruptionToOpportunity}. 
%However, prior evidence suggests that automation can reduce employment and wages when deployed as a substitute for routine labor rather than a complement to skilled work~\citep{AcemogluRestrepo2020Robots}. Because the benefits of new technologies are not shared automatically, workforce considerations must play a central role in how these tools are developed and deployed~\citep{AcemogluJohnson2023Power,AcemogluAutorJohnson2023ProWorker}.

%Since technological advancement does not improve all lives equally, it is critical to include a workforce perspective in the conversation of adopting new technologies to meet workforce needs~\cite{AcemogluJohnson2023Power, AcemogluAutorJohnson2023ProWorker}.

%Technological change is shaped by economic incentives and institutional environments; technologies can be directed toward labor-replacing or labor-augmenting applications depending on how they are developed and deployed~\citep{Acemoglu2002DirectedTechnicalChange}.

\paragraph{AI as a Rising Solution}

Artificial intelligence (AI) is emerging as a key solution to growing efficiency and output demands in manufacturing. Industry reports consistently identify AI as a critical enabler for maintaining productivity amid workforce shortages and rising system complexity. The \textit{World Economic Forum} emphasizes that AI-driven tools and digital technologies are essential for augmenting the industrial workforce, addressing attrition and skills gaps through intelligent automation, data-driven training, and continuous upskilling~\citep{WEF2025EmpoweringFrontlines}. The \textit{IEEE Technology Megatrends 2024} report likewise positions AI and automation, particularly generative and agentic systems, as central to sustaining efficiency across digital transformation and sustainability initiatives, noting that AI now underpins nearly all emerging industrial technologies~\citep{ieee2024}. Xometry and Thomas’s \textit{2026 Manufacturing Trends Report} identifies ``AI as a Competitive Necessity'' as its leading insight, with 82\% of manufacturers citing AI as a primary driver of growth and 44\% reporting significant returns on investment from AI adoption~\citep{xometry2026}. Yet a persistent skills gap in the use of AI tools underscores the need for targeted upskilling to fully realize these benefits. It is therefore critical to understand how AI currently integrates, and will continue to evolve, within engineering and manufacturing workflows, and to establish the systems and training necessary to achieve its positive potential.

\subsection{Artificial Intelligence and Agentic Systems}

%Define common terms like AI, ML, generative AI, LLMs, and agentic AI
AI refers to the ability of a computer system to perform tasks that mimic human intelligence such as learning, reasoning, and problem solving. Machine learning is a subset of AI that utilizes algorithms to learn from data~\citep{MicrosoftAzure2025}, and deep learning is a subset of machine learning that uses neural networks with many layers to learn complex patterns. A major breakthrough in deep learning came in 2017, when Vaswani et al. introduced the transformer~\citep{vaswani2017attention}, a neural network architecture based on the multi-head attention mechanism. This innovation fundamentally reshaped natural language processing and paved the way for large language models (LLMs), which leverage large-scale transformer architectures to learn from massive corpora and support a wide range of general-purpose reasoning and generation tasks.

\paragraph{Agentic AI}
Building on recent advances in LLMs and multimodal or vision-language models (VLMs), a new class of \textbf{agentic AI systems} has emerged. Unlike traditional models that passively generate outputs, agentic systems can pursue goals, plan multi-step actions, interact with external tools and software, and adapt based on feedback~\citep{IBM2024AgenticAI}. While there is no universally agreed-upon definition of an “agent,” we adopt a definition closely aligned with Anthropic’s: “an AI system equipped with tools that allow it to take actions, like running code, calling external APIs, and sending messages to other agents”~\citep{anthropic2024measuring}. This definition is also consistent with Russell and Norvig’s broader characterization of an agent as “anything that can be viewed as perceiving its environment through sensors and acting upon that environment through effectors”~\citep{10.5555/773294}.

We distinguish agentic systems from both traditional workflow automation and LLM-based copilots. Workflow automation executes predefined, deterministic pipelines without dynamic reasoning. LLM copilots provide human-initiated suggestions but do not independently orchestrate multi-step tool usage. Agentic systems, in contrast, can select and sequence tools in pursuit of a defined objective. Within engineering and manufacturing workflows, this may involve interacting with computer-aided design (CAD), computer-aided engineering (CAE), computer-aided manufacturing (CAM), enterprise resource planning (ERP) systems, simulation tools, or internal databases.

However, our empirical findings indicate that most companies are not currently deploying fully autonomous agents. Instead, current implementations range from predefined automation to forms of \textit{bounded autonomy}, where agents operate within tightly scoped workflows, remain subject to human validation, and do not independently assume engineering accountability. In fact, a 2026 report by Anthropic, a leading AI research company, indicates that the observed AI coverage is much lower than the theoretical AI coverage possible in engineering~\citep{massenkoffmccrory2026labor}. This observation aligns with the task landscape described later in this paper.

Accordingly, our interviews span both existing non-agentic AI deployments and forward-looking visions of fully agentic systems. Our analysis focuses primarily on this intermediate regime of semi-autonomous, tool-using agents embedded within established engineering processes.

\subsection{Objectives and Scope of This Work}
The goal of this work is to understand how AI, and in particular agentic systems, will impact engineering and manufacturing workflows from an industry perspective. We interviewed key players in this domain, including large engineering enterprises, small/medium engineering enterprises, manufacturers, CAD/CAM/CAE tool providers, and AI developers. These interviews allowed us to gather diverse perspectives on how AI is currently being used, how it could be integrated into future workflows, and what barriers and enablers govern its adoption.

Given prior evidence that the effects of automation depend critically on how technologies are deployed and whether they complement or substitute for labor~\citep{AcemogluRestrepo2020Robots,AcemogluAutorJohnson2023ProWorker}, our analysis intentionally centers the perspectives of engineers, operators, executives, and tool builders directly shaping AI integration within their organizations.

From these conversations, we identify the most promising opportunities for AI to augment engineering work, the challenges that limit its utility, and the broader trends influencing organizational readiness and trust. Our aim is to equip both AI developers and engineering/manufacturing organizations with a clear picture of where AI and more specifically agentic systems can deliver value in practice, what barriers constrain adoption, and what capabilities and IT infrastructure are needed to support more advanced applications. In addition, insights from CAD, CAM, and CAE tool providers highlight best practices for designing effective software tools and reveal how industry leaders are beginning to embed AI and, increasingly, agentic capabilities into their products.
% \todo[inline]{Perhaps make these the actual findings rather than an explanation of what we did. Such as the general findings of the task landscape, the barriers and needed breakthroughs, trust and governance, and adoption dynamics.}

This paper makes three contributions:

\begin{itemize}
    \item Synthesizes industry perspectives from more than 30 in-depth interviews spanning large enterprises, SMEs, AI developers, and CAD/CAM/CAE developers to characterize how AI and agentic systems are understood and adopted in practice. 

\item Identifies the engineering task landscape for AI, highlighting near-term, well-posed opportunities (repetitive and data-intensive work) versus higher-stakes applications where verification, trust, and governance requirements dominate. 

    \item Surfaces practical barriers and enabling breakthroughs (data accessibility, representation, legacy tool integration, security, privacy constraints, spatial reasoning needs, verification, governance mechanisms) that stakeholders believe will unlock broader utility and adoption.
\end{itemize}

\begin{figure}
    \centering
    \includegraphics[width=0.9\linewidth]{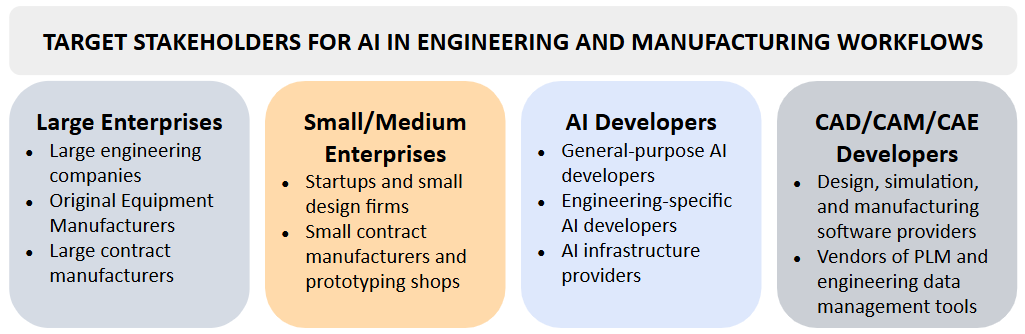}
    \caption{We performed 33 interviews with members across these categories.}
    \label{fig:stakeholders}
\end{figure}

\section{Methodology}
We conducted 33 interviews (across 28 companies) with leading experts from the four categories outlined in Figure~\ref{fig:stakeholders}: large engineering or manufacturing enterprises, their small or medium counterparts, AI developers, and CAD/CAM/CAE developers. These included CEOs, CTOs, and founders of large enterprises, startups, and contract manufacturers. It also included the heads of AI at many large engineering firms, as well as junior and senior employees across these four categories whose work streams would be directly impacted by the adoption of AI. 

\paragraph{Who we Interviewed:} The number of companies interviewed across the four stakeholder categories was as follows: large enterprises (6), small and medium enterprises (14), AI developers (5), and CAD/CAM/CAE developers (3) as shown in Table~\ref{tab:interview_companies_long}. Participants were recruited using a combination of purposive sampling and referral-based (snowball) sampling. Within each of the four stakeholder categories, we first identified organizations that were actively engaged in engineering design, manufacturing operations, or AI system development. Selection criteria included leadership roles in engineering or technical decision-making, direct involvement in workflows likely to be impacted by AI adoption, or demonstrated activity in AI experimentation or deployment.

Initial outreach occurred via direct email invitations to executives, technical leaders, or domain experts identified through professional networks, conferences, and industry partnerships. In several cases, interviewees referred us to additional colleagues or peer organizations with relevant expertise, allowing us to expand the sample through snowball sampling. This approach enabled access to individuals with firsthand implementation experience and practical insight into adoption dynamics. 

\paragraph{How we Interviewed:} All interviews followed a semi-structured format guided by a common interview protocol (Appendix, Section~\ref{sec: interview_questions}). Interviews typically lasted 30–60 minutes, with shorter (10–15 minute) interviews conducted in conference settings. The interview guide included open-ended questions covering current understanding of agentic AI, current adoption practices, and barriers around using AI in general, and what it takes as an individual and an organization to trust AI in their workflow. We also probed about the most exciting use cases, the most promising ones, and the highest impact ones. Although the structure was consistent, the follow-up questions were tailored to each stakeholder’s domain to investigate sector-specific constraints and opportunities. %Through our interviews, we uncovered key elements of AI adoption and utility, which are explored individually in Sections~\ref{sec:task_landscape}-\ref{sec: trust-governance}, these are the task landscape, the state of engineering data, the current barriers and needed breakthroughs for AI adoption, trust and governance around AI. Finally in Section~\ref{sec: adoption-dynamics}, we synthesize the dynamics for both current and future adoption, including how it varies across and within organizations, and what prerequisites are required for full adoption. 

Each interview was conducted by at least two members of our research team, with one person serving as the lead interviewer. Most interviews were held in person or via video call, with one exception in which a small/medium enterprise CTO provided written responses by email. At the start of every interview, we informed participants that their responses would remain confidential unless they granted explicit permission to attribute information to their company. This measure was intended to encourage openness and candor.

\paragraph{Analysis of Interviews:} All interviews carried out by video call or in person were recorded with participant consent and then automatically transcribed using Otter or Zoom transcription services. We analyzed the interviews using an iterative and inductive approach, meaning that themes were derived from the interview data itself rather than from a predefined framework. Transcripts were read in full to identify recurring concepts such as task types, data constraints, and organizational dynamics. Common concepts were grouped into higher-level themes, with themes elevated when they appeared across multiple stakeholder categories rather than within a single organization. 

Analysis was conducted collaboratively. When disagreements arose, transcripts were revisited and discussed until consensus was reached. To reduce confirmation bias, we included both AI-enthusiastic and AI-skeptical organizations and examined cases of non-adoption alongside successful deployments.

We note that this is a qualitative study and the findings reflect the perspectives of the interviewed sample, and should be interpreted in light of the study’s limited size and non-random sampling. The following sections draw on these interviews to characterize the task landscape, barriers, and enabling conditions shaping AI adoption in engineering and manufacturing workflows.

%\todo[inline]{Include the mix of roles of interviewees. Provide a 1-2 sentence description of each company mentioned by name in the manuscript to provide context for the audience.}

%specifically defining the utility of AI adoption as a function of baseline tasks, higher value tasks, and the trust,

\begin{longtable}{|p{3cm}|p{12cm}|}
\caption{Interviewed Companies and Organizational Descriptions}
\label{tab:interview_companies_long} \\
\hline
\textbf{Company} & \textbf{Description} \\
\hline
\endfirsthead

\hline
\textbf{Company} & \textbf{Description} \\
\hline
\endhead

\hline
\multicolumn{2}{r}{\textit{Continued on next page}} \\
\endfoot

\hline
\endlastfoot

\rowcolor{blue!5}
\multicolumn{2}{|c|}{\textbf{Large Engineering or Manufacturing Enterprises}} \\
\hline

National Aeronautics and Space Administration (NASA) &
United States government agency responsible for civilian space exploration, aeronautics research, and space science, conducting large-scale engineering and research programs. \\
\hline

GE Vernova &
Energy-focused industrial company providing technologies and services for power generation, grid infrastructure, and energy transition systems. \\
\hline

Anon. & %Analog Devices
Semiconductor company designing and manufacturing analog, mixed-signal, and power management integrated circuits for industrial, automotive, and communications markets. \\
\hline

Anon. & %L3H
Major defense and aerospace contractor designing and manufacturing advanced systems for communications, sensing, and national security applications. \\
\hline

Anon. & %Aerospace Corp
Federally funded research and development center providing technical analysis, engineering support, and systems assessment for national security space programs. \\
\hline

Anon. & %Anduril
Defense technology company developing autonomous systems, sensors, and software-enabled hardware platforms for military and security applications. \\
\hline

\rowcolor{blue!5}
\multicolumn{2}{|c|}{\textbf{Small/Medium Engineering or Manufacturing Enterprises}} \\
\hline

Inkbit &
Additive manufacturing company specializing in multi-material 3D printing technologies for complex and functional parts. \\
\hline

Cyvl &
Infrastructure technology company providing roadway inspection and asset mapping solutions using sensor data, computer vision, and analytics for public-sector customers. \\
\hline

Alloy Specialties &
Precision manufacturing and supply chain services company specializing in complex machined components and assemblies for aerospace and defense, including turbine disks and rotor parts. \\
%Engineering and manufacturing services company supporting product development, production, and supply-chain workflows for industrial clients. \\
\hline

Saunders Electronics &
Contract manufacturer specializing in printed circuit board assembly (PCBA) and electromechanical system integration. \\
\hline

Anon. & %Multiply Labs
Robotic manufacturing company developing automated systems for pharmaceutical compounding and medication preparation. \\
\hline

Anon. & %Gecko
Robotics and inspection company using autonomous systems and sensor data to assess industrial infrastructure and create digital representations of physical assets. \\
\hline

Anon. & %Chaos Industries 
Defense technology company focused on developing advanced systems for national security applications, including sensing and signal-processing technologies. \\
\hline

Anon.  & %JRC
Defense and aerospace contractor providing systems engineering, strategic weapons development, and radiation-hardened microelectronics for national security and deterrent programs.. \\
\hline

Anon. & %Forterra
Defense and commercial robotics company developing autonomous systems and solutions for military and civil engineering applications. \\
\hline

Anon. & % Reach Space Technologies
Space technology company developing commercial rocket propulsion. \\
\hline

Anon. & % Janco Electronics
Electronics manufacturing services provider offering PCB assembly and related production services. \\
\hline

Anon. & % SVTronics
Electronics contract manufacturer specializing in printed circuit board assembly and systems integration. \\
\hline

Anon. & %Electronic Interconnect Corp 
Printed circuit board manufacturer offering PCB fabrication and assembly services. \\
\hline

Anon. & %Printed Circuit Corp
Manufacturer of printed circuit boards serving aerospace, defense, and industrial customers. \\
\hline

\rowcolor{blue!5}
\multicolumn{2}{|c|}{\textbf{AI Developers}} \\
\hline

LangChain &
Developer of open-source software frameworks and commercial platforms for building applications with large language models, including agent orchestration, tool integration, and observability. \\
\hline

C3 AI &
Enterprise software company providing AI applications and development platforms for industries such as energy, defense, manufacturing, and logistics. \\
\hline

Synera &
Software company providing an automation platform for mechanical engineering workflows, enabling integration of CAD, CAE, and simulation tools through low-code and API-based approaches. \\
\hline

AllSpice &
Provider of collaboration and design review software for electrical engineering teams, supporting schematic and PCB design workflows with automated checks and review tooling. \\
\hline

Anon. &  %Liquid AI
Artificial intelligence company developing foundation models designed for efficient deployment, including on-device and edge environments. \\
\hline

\rowcolor{blue!5}
\multicolumn{2}{|c|}{\textbf{CAD/CAM/CAE Developers}} \\
\hline
Autodesk &
Developer of software products and services for architecture, engineering, construction, manufacturing, media and other industries. Examples of their 3D design software include AutoCAD and Autodesk Inventor. \\
\hline

Parametric Technology Corporation (PTC) &
Developer of product lifecycle management (PLM), CAD, and industrial software tools used to manage product design, engineering, and manufacturing processes. \\
\hline

Siemens &
Global technology company operating across industrial automation, digital engineering software, energy systems, and manufacturing technologies. \\
\hline

\end{longtable}

% \section{Elements of AI Adoption and Utility}
% \label{sec:elements}

\section{The Task Landscape for Agentic AI Adoption}
\label{sec:task_landscape}
Drawing on the interview data, our findings suggest which engineering tasks are well-posed for current AI capabilities and which remain challenging. Engineers consistently identified repetitive, high scale tasks and data-intensive work as the most viable near-term opportunities. Beyond initial adoption, our interviewees described growing value in process orchestration and multi-agent coordination tasks, as opposed to single-task automation. In safety-critical contexts, AI remains primarily an advisory tool—offering suggestions that humans must validate rather than enabling fully autonomous decision-making. Looking ahead, engineers anticipate new interaction paradigms that could further expand AI’s agency and utility, although significant infrastructure constraints must first be addressed. Below, we outline the task landscape surrounding AI adoption in engineering.
\subsection*{Key Takeaways:}
\begin{itemize}
    \item \textbf{Repetitive, high-volume tasks with clear structure} represent near-term AI adoption opportunities.
    \item \textbf{Data-intensive work where information is large scale, inaccessible, or difficult to synthesize} also show direct AI opportunities - with some engineers spending 25\% of working hours on this area alone.
    \item \textbf{Process orchestration and multi-agent coordination} are emerging as high-value use cases for AI, beyond single-task automation.
    \item \textbf{Safety-critical applications still use AI in an assistive role,} providing suggestions for humans to validate rather than making autonomous decisions.
    \item \textbf{Engineering trust in AI} varies widely across industries.
    \item \textbf{New interaction paradigms} are limited by missing infrastructure.
\end{itemize}
\subsection{Well-posed Tasks}
\label{subsubsec:well-posed_tasks}
From our interviews, three categories of engineering work emerged that are well-suited for AI, including structured, repetitive, and high-volume activities; data-intensive tasks with processing or synthesis constraints; and multi-step coordination workflows. These well-posed tasks share clear success criteria and established patterns, as discussed below.

\paragraph{1. Repetitive and High-Volume Activities}

One type of engineering task involves working with clearly structured data at high volumes. For example, in integration engineering, requirements processing requires digesting massive documents for systematic analysis. 
\begin{myquotebox}
Requirements documents can be from 50 to 1000 pages long. The requirements are then compared in a matrix and then guidance for how mechanical will impact electrical design and battery management [is developed], all the way into software.
\attributed{An engineering manager at a major defense contractor.}
\end{myquotebox}
% An engineering manager at a major defense contractor attested to this stating “[r]equirements documents can be from 50 to 1000 pages long. The requirements are then compared in a matrix and then guidance for how mechanical will impact electrical design and battery management [is developed], all the way into software.” 
Beyond basic impact suggestions, this engineer notes that AI could “predictably guide the design… predicting a list of problems and risks to design efforts.” This highly patterned task involves synthesizing requirements, building traceability matrices, and propagating cross-functional impacts, and thus is a prime opportunity for AI augmentation. From a research perspective, works such as~\citep{edwards2024advise, Marshall2019, Edwards2022NLPFiltering, 10.1093/jamia/ocaf063} have demonstrated the use of language models to filter through large corpora of documents. 

Similarly, in the manufacturing context, one major repetitive and high volume task is ingesting diverse customer data. In defense manufacturing, suppliers often work with multiple prime contractors simultaneously and have found they each have unique quality clauses, acceptance criteria, and requirements documentation. This involves extracting key inspection points and critical dimensions and then filling out standard paperwork and templates. Defense contractors highlight how customers have unique requirements and expectations that must be understood and met. This data sorting and entry task is clearly structured and high volume, and thus has been found to be well-suited for AI augmentation.

This repetitive data entry problem is even faced when working with just one customer. For example, a manufacturing engineer at a small electronics manufacturing company reported that ``the tedious stuff I find here is there seems to be a lot of data entry, a lot of taking repeat information and putting it in our system.'' Specifically, they noted that these tedious tasks include ``putting in the descriptions [of components] into [their ERP system]. The other thing I would say is definitely comparing the old rev to the new revs, and seeing what has changed, because that can really ... become an issue if we don't catch it.''

In fact, four of the five contract manufacturing firms we spoke to explicitly mentioned creating ``in-house'' company tools (like software scripts) to automate certain repetitive tasks. A manufacturing engineer at Saunders Electronics noted that automatic optimization for machine programming already delivers meaningful time savings and is trusted in day-to-day production. That said, some mentioned that these tools have been difficult to scale or maintain for reasons such as ``the person who created the software quit several years ago… he didn’t leave a lot of notes in his code,'' and the infeasibility of hiring a full-time in-house software engineer. This indicates a need for more standardized and scalable automation tools. For certain repetitive tasks in electronics manufacturing, such as quality inspection, automation tools like automated optical inspection (AOI) which utilizes computer vision have been developed and are used in practice~\citep{9214824}. 

In architectural design workflows, similar repetitive tasks exist within parametric contexts. As designs mature, there are frequent modifications within physics and regulatory constraints that must preserve overall requirements. A senior research and design engineer at Autodesk highlighted, “[i]terative tasks like ‘making the building longer or shorter or add[ing] windows.’ Instead of painstakingly doing that manually, leveraging an agent to optimize but still be physics informed (is the structure stable, is there sufficient sunlight).” Leveraging AI in such parametric and constrained environments will allow faster iteration as it can propose multiple options that meet all requirements.

Software engineering has equivalently repetitive tasks such as navigating undocumented codebases. A software engineer from a startup highlighted how they “spend so much time [since their] code base is extremely chaotic… it’d be so nice to be able to [search] ‘where does this variable come from?'” Such tasks require tracing through call stacks and developing a mental model of how data flows through a module. This finding is not an outlier, as the CTO at Inkbit identified “documentation and identifying root cause[s] of issues” as a time consuming task that is ripe for augmentation. AI can perform the tedious tasks of parsing through code structures, develop call graphs, trace data flow, perform root cause analysis, and document failure modes - allowing for more engineering productivity. This view is of alleviating tedious tasks is corroborated by the findings of Armstrong et al. who found that people performing ``complex problem-solving tasks and new idea generation tasks are the best predictors of workers championing new technologies'' like AI automation~\citep{Armstrong2024AutomationWorkers}. Likewise, prior research shows that ``by automating routine tasks, AI frees up engineer's time, allowing them to focus on creative problem-solving, ideation, and innovation''~\citep{chinda2023}.

In the infrastructure inspection field, there are prime examples of automation opportunity. An engineer at Cyvl highlighted how government agencies use “pen and paper… [and] Excel” for road defect tracking and how redundancy exists since “[Multiple road crews] will go out and look at [the same] … pothole” just to confirm its presence. The Cyvl engineer emphasized how this tedious and expensive process could be augmented using road data collection equipment to identify cracks and potholes. The collected data could then be automatically formatted into structured reports, enabling inspection crews to work more efficiently.

\paragraph{2. Data-Intensive Work} 

Another type of well-posed tasks are those which have clear objectives but are limited by data access, volume, or synthesis. An engineer at NASA highlighted how in early AI adoption, “[they] see a lot of value in making historical data readily available. [They] have all this data across [their] many missions, but it’s not easy to access.” For organizations like NASA, leveraging information from decades of missions such as past designs, test results, anomaly reports, and lessons learned can yield much higher effective engineering productivity although is frequently inaccessible or difficult to digest - indicating a great opportunity for utilizing the organization capabilities of AI. Likewise, comparing similar historical parts can allow for process and cost comparison. This task requires a vast knowledge of existing designs and would otherwise be extremely time consuming to derive actionable information. While geometric comparison is still in its infancy, some companies see great value in using AI to sort components into product families as it can allow for scale through process design.

Furthermore, routine documentation work commonly consumes a large portion of engineering time. When asked what he would prefer his engineers work on, the general manager of an electronics manufacturer, noted that manufacturing engineers create the most value when they are making decisions and resolving issues, rather than spending time on repetitive tasks such as data entry. An engineer at NASA highlighted how for a sample project “writing and reviewing documents, checking compliance, generating reports … [took up] a quarter of [their] time.” Some engineers liken such tedious tasks to administrative overhead rather than value-add engineering work. Thus such a data-intensive process indicates a clear opportunity for AI augmentation as it can draft, check, and generate routing documentation and accelerate productivity. Early stage work by NASA JPL showed a proof-of-concept for using AI for otherwise time-intensive document generation, specifically generating instructions for build, assembly, and test of new hardware. While their work explored a small scale of problems (just three examples), it revealed potential in this space~\citep{Nuernberger2024AiBAT}. 

Throughout the design process, each step can be iterative, time consuming, and coupled--requiring overhauls for each change. 
% An engineer at NASA emphasized this issue: “[starting with] CAD… you sketch, extrude… then the simulation part… you turn it into a mesh… the thermal person turns it into a different mesh… people have spent a week making a mesh so you don’t want to change anything.” 
\begin{myquotebox}
[Starting with] CAD: you sketch, extrude; then the simulation part: you turn it into a mesh, the thermal person turns it into a different mesh. People have spent a week making a mesh so you don’t want to change anything.
\attributed{A NASA engineer.}
\end{myquotebox}
This cross-functional pipeline can take weeks for complex geometry, which introduces reluctance to iterate designs. Such intensive processes present opportunities for AI to enable rapid design exploration through automated pre-processing, like mesh generation, and post-processing, like result extraction.

In environments with sensor-rich systems, the volumes of data generated create challenges with sifting through readings and discerning which signals are reliable enough to be customer-facing. An engineer at a robotics start-up stated that “data validation is probably ripe [for AI] … because it’s so data dense … [and] the readings are not perfect… You have to figure out which readings you actually want to present to the customer.” In many cases, engineering judgement is required to determine reliable signals to avoid losing customer trust. Although, since this task is systematic, there is a great opportunity for AI to be leveraged for identifying anomalies and learning the normal ranges of data.

A similar pattern appears in engineering release workflows. Before a design is approved for manufacturing, drawings must pass through a formal release process to ensure they are complete and compliant with internal standards. An engineering manager at a major defense contractor described how engineers spend significant time validating drawings prior to release, ``mak[ing] sure the drawings are clean and correct.” While this task follows a straightforward process, execution is tedious and intensive when checking every drawing and every dimension. Thus, this task is ripe for automation, since the validation rules are well-defined.

Electronics design review follows a similar pattern of data-intensive checking. It is currently a very manual process that involves checking circuit designs across datasheets, checking component specs, and flagging mismatches. AllSpice are directly working to augment these workflows with AI agents “that will … go check all of your designs, cross-reference against manufacturing data sheets, and … alert you if there are any errors in proposals” as a “first pass for the engineering teams,” as highlighted by an executive there. Such well-defined but time-consuming tasks like retrieving datasheets, extracting specifications, and then preliminary comparisons is prime for AI automation.

When adapting models to reflect test data, engineers follow a labor-intensive manual process of tuning model parameters and iterating dynamic models to fit to the breadth of data collected. The director of structural dynamics and environments for a large aerospace company reflected how this workflow is “usually a manual process” and how he has “tried to use automated versions … to create at least a starting point.” This concept of automating the preliminary work is common throughout industries and reflects a common theme of how early AI implementation focuses on augmenting work rather than replacing it.

\paragraph{3. Process Orchestration}

Process orchestration augmentation is highlighted by innovation in supplier-OEM interactions. An engineer working at a company developing AI highlights how multi-step request for quotation (RFQ) workflows (the process of an OEM sending requests, followed by the supplier configuring parts, performing manufacturability analysis and cost calculations, followed by responding to the OEM) are becoming orchestrated by agents and represents a highly beneficial area for automation. The engineer explained how some companies have highly iterative processes that follow a similar workflow for each project and thus “want to capture this information.” AI automation enables these complex patterns to be captured and leveraged to isolate findings and successful approaches.

Teams of AI agents are emerging as an alternative to single-agent models. A software engineer at LangChain mentioned how ``[a] lot of [their] customers’ use case is around internal tooling… especially at PE firms, consulting'' and how ``case analysis and initial research can be orchestrated amongst a team of agents''. In this sense, complex tasks are split across the agent team, with each agent working as a specialist in parallel, and an orchestrator agent working to integrate their results. They emphasize that defining the interactions and workflows is more developer-facing upfront although allows for a sophisticated, repeatable workflow. 

\subsection{Challenging Tasks for the Future}
\label{sec:Future_challenging_tasks}
As speculated by an engineer at NASA, ``As things go on, and agents become good, and the tools that they have access to become good… [we may] eventually have … nearly … full-stack automation of development''. At present, many challenges exist before full-stack dependency will exist. Two main areas of focus are high-stake, safety-critical applications and new interaction paradigms:

\paragraph{1. High-Stakes, Safety-Critical Applications}
In contexts where safety is paramount, AI is used in an assistive role rather than an autonomous decision-maker. With the current technical capability of cutting-edge AI, engineers at large aerospace companies use it to generate solution candidates for humans to then validate. The director of structural dynamics and environments for a large aerospace company highlighted how AI is used for augmentation rather than automation by providing suggestions, since humans still apply their judgment and domain expertise. This augmentation reflects the current capability limits of AI and how some organizations are not comfortable with full reliance on AI in safety-critical workflows. Similarly, in validation work, AI automation is applied for checking, comparing, and initial screening to flag potential issues, although the final acceptance decision is made by experienced human engineers. This trend is summarized succinctly by Chinda and Gin: ``Effective human-machine collaboration requires clear communication, trust, and shared decision-making. Engineers must understand the capabilities and limitations of AI systems to effectively leverage their potential''~\citep{chinda2023}. In the case of safety-critical applications, understanding when to leverage AI and when to apply engineering fundamentals is an important engineering decision.

In high stakes scenarios, trust in AI systems is built gradually through demonstrated reliability. This trust is developed through accumulation of evidence across many use cases. Interpretability and hallucination are major concerns for AI outputs since they commonly are black-box decisions, without insight into reasoning. As a result, the attitudes of interviewees across fields vary dramatically. One engineer from a robotics start-up was skeptical about using a AI programming guidance because they found they frequently had to rewrite the output code. At the same time, some organizations believe existing review processes are well-suited by AI outputs. In an aerospace context, throughout the requirements review, design review, and test readiness review, teams and subsystems review each other’s work to provide checks and balances. Across large aerospace companies like NASA, some engineers believe the existing checkpoint systems might integrate well with AI - suggesting that humans can make mistakes just like AI, and thus the checks in place should hopefully catch those mistakes.

\paragraph{2. New Interaction Paradigms}
As AI advances, organizations are beginning to imagine new interaction modalities. For example, a software engineer from LangChain discussed ambient agents, which are “systems that remain dormant until activated by specific triggers … LangChain’s graph-based architecture enables the coordination of multiple such agents in sequence to accomplish complex, event-driven tasks.” Once such agents are configured, instead of waiting for the next prompt, they run automatically and respond to events.

With new advancements, additional terminology distinctions are emerging, such as a co-pilot vs and agent. A software engineering from LangChain explains, “co-pilots function primarily as human-facing conversational assistants … [while] agents extend beyond dialogue to perform autonomous, tool-integrated workflows … without continuous human prompting.” With these two different interaction paradigms, different uses cases become appropriate for each. In an engineering context, organizations are finding that they can benefit from both, co-pilots for exploration and agents for automation and orchestration.

As a long-term goal, some engineers envision an immersive collaborative AI in shared virtual workspaces, with multiple engineers participating in real-time. As one engineer at NASA put it, “ultimately, success looks like multiplayer Jarvis [from Iron Man], where… [myself and my colleagues] put on some kind of XR headset, and we have AI agents collaborating with us … building [a design], … moving things around, … and it’s giving us performance predictions.” A major aspect of this vision is rapid iteration and multiple design options, which was found to be a common sentiment among organizations when forecasting the major engineering benefits of AI.

Such future interaction modalities require infrastructure changes that are not prevalent today. As highlighted by a NASA engineer, “digital employees need digital tools.” In order for AI agents to interact with manufacturing suppliers, component vendors, or machine shops, many of them need programmatic interfaces. That same engineer reiterates, ``there’s no API for machine shops. There’s no API that we can access for most suppliers when performing component selection.'' Most of these ecosystems are designed for human interaction, not programmatic access. Moreover, technical details and cost figures can be hidden behind sales channels for more advanced components, which only worsens the effectiveness of agentic solutions that need to access that data.

\section{The Current State of Engineering Data}
\label{sec:State_of_eng_dta}
Our interviews revealed that data is the major constraint on AI adoption in engineering organizations. More specifically, the availability, accessibility, and quality of engineering data. Across aerospace, defense, manufacturing, and hardware development, companies indicated that data challenges were consistently more limiting than algorithmic challenges. The main aspects of these challenges include security and regulatory constraints that prevented cloud-based AI adoption, data fragmentation across disconnected systems and formats, and essential knowledge being un-codified. 
% The Chief Executive Officer of Alloy Specialties emphasized, ``it's not the [AI] tools that are the problem. It's the dataset.''
\begin{myquotebox}
It's not the [AI] tools that are the problem. It's the dataset.
\attributed{CEO of Alloy Specialties.}
\end{myquotebox}
A leader at PTC emphasized the data gap ``Few of the companies we work with have enough data to create their own foundational model, which makes aligning their specs with what LLMs understand a real challenge.'' Below, the key data challenges surrounding AI adoption in engineering are listed.

\subsection*{Key Takeaways:}
\begin{itemize}
    \item \textbf{Security, IP protection, and regulatory compliance (ITAR, GDPR, ISO)} prevent access to state-of-the-art cloud-based AI capabilities.
    \item \textbf{Engineering data is scattered across various sources} within organizations - causing retrieval to take weeks or months, especially with workforce retirement.
    \item \textbf{Critical information exists in various machine-unfriendly formats} that require specialized parsing before they can be actioned by AI.
    \item \textbf{Current AI systems struggle with robust 3D geometric understanding and CAD-to-CAM translation} due to the spatial reasoning capability needed for manufacturing automation.
    \item \textbf{Most valuable trade knowledge is kept in expert's heads} rather than being codified, which threatens knowledge loss. Captured knowledge can become stale as materials and methods evolve.
\end{itemize}

\subsection{Privacy / IP / ITAR} 
\label{sec: data-privacy}
Strict security, IP, and regulatory constraints are imposed on a lot of high-value engineering data such as defense systems, aerospace designs, and proprietary manufacturing processes. In many instances, the data must be physically isolated from cloud infrastructure through multiple mechanisms, such as air-gapped networks, removable drives locked in physical safes, and strictly controlled information tunnels. When handling sensitive data, service providers often are forced to use country-specific data centers to comply with the data sovereignty laws of their client’s country. Additionally, large aerospace companies struggle with leveraging AI connected to the internet due to security concerns, as further discussed in section~\ref{sec: security-privacy-deployment}. As a result of these constraints, sensitive engineering data often cannot be given to high-capability cloud-based AI models, thus limiting the amount of data available for training and use by AI systems. The architectural implications of these security requirements are further discussed in \ref{subsec:prereq_adoption}.

\subsection{Data Fragmentation}
\label{sec:Data_fragmentation}
 %As emphasized by one engineer leading an AI initiative in their company, they ``interviewed 22 of our mission development subject matter experts… [and] found 26 different data sources.''  This finding is common across many engineering organizations and underscores a huge challenge for AI integration.

A common challenge throughout different industries is that engineering data is scattered. Many of the engineers interviewed mentioned that no single integrated system exists in practice and that, instead, data is scattered across different storage types: personal hard drives, SharePoint sites, custom databases, and ad-hoc file shares. This scattering poses significant challenges when there are multiple instances of data, different versions, and non-standard naming conventions that all cause a lack of interoperability and usability. 

\begin{myquotebox}
    I interviewed 22 of our mission development subject matter experts...  and found 26 different data sources.
    \attributed{An engineer at a large enterprise.}
\end{myquotebox}
The impacts on day-to-day engineering work can be severe. As highlighted by one engineer at a large enterprise, when attempting to find historical designs, ``finding that data could take you weeks or months of back and forth,'' indicating that ``you [often] have to find the person that [designed] it to find the data''. A study by the McKinsey Global Institute found that workers can devote 30 to 40 percent of their time searching for data if data storage is disorganized~\citep{grande2020reducing_data_costs}. 
\begin{figure}
    \centering
    \includegraphics[width=0.9\linewidth]{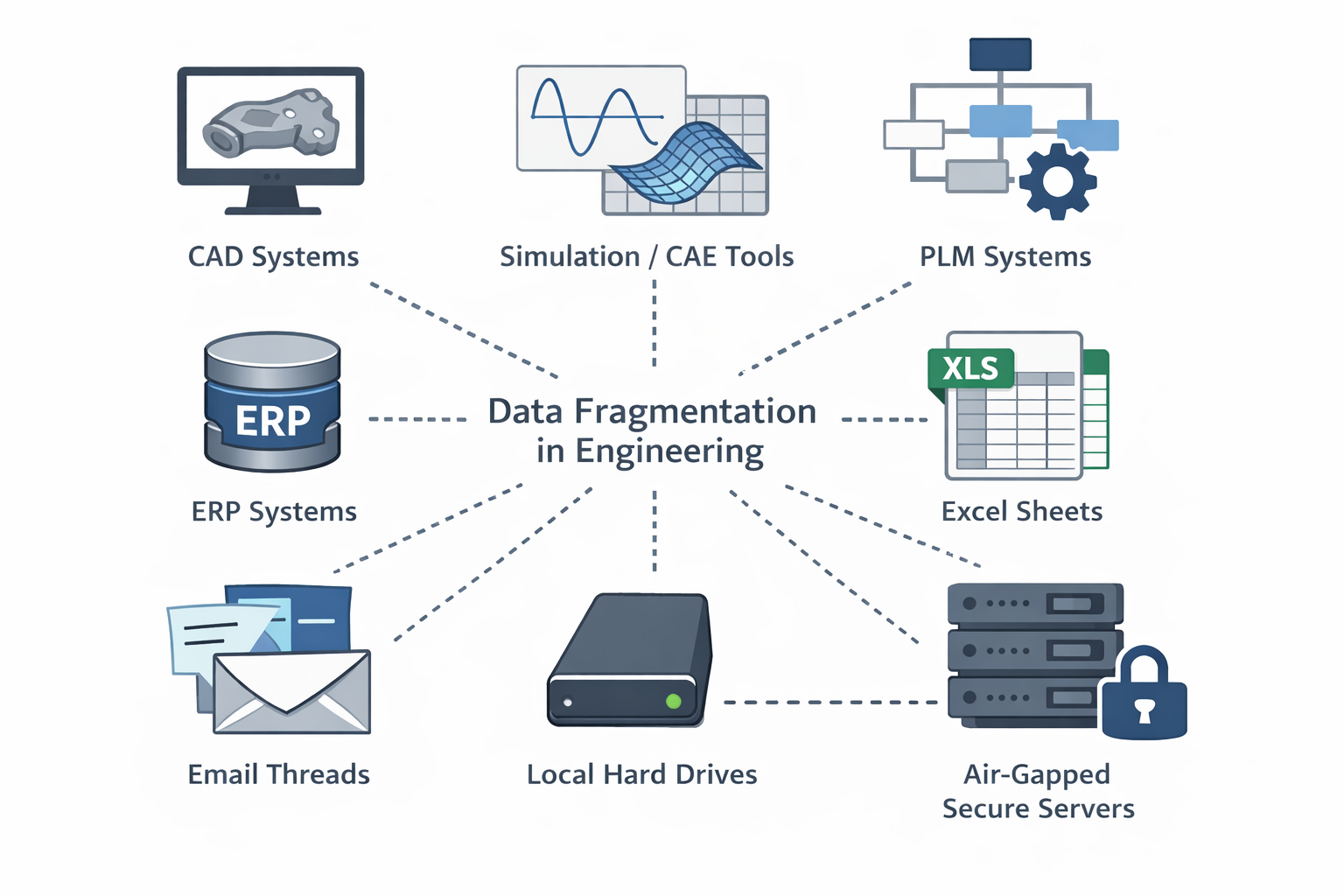}
    \caption{Examples of the various sources of data in engineering and manufacturing workflows.}
    \label{fig: fragmentation}
\end{figure}

The task of locating data becomes especially challenging when people transfer between teams, between divisions, or leave the company. Without that person’s knowledge, data effectively becomes lost despite it existing somewhere. One engineer even referred to this process as ‘people archaeology’. Workforce turnover and key staff retirement only exacerbate this problem as their mental map of where data is stored disappears. As reiterated by the CEO of Alloy Specialties, ``we’re on the precipice of a mass retirement” since the “average age in [small manufacturing shops] is about 55 from an employee perspective, and the average age of ownership is in their 60s''. They also emphasized how this will cause not only a loss of skilled talent but also a loss of precision manufacturing knowhow which is not codified

Larger organizations face additional structural fragmentation as different business units tend to maintain separate systems. Each business unit can have different naming conventions, standards, and data schemas, all of which were not designed to be interoperable. Despite integration efforts, the reality is that data format and structure harmonization can take years. When attempting to consolidate engineer data, the CEO of Alloy Specialties mentioned how ``[t]he major problem is that datasets aren’t clean'' and that [the] next major issue in all industrial companies is that your datasets are completely disparate''. This can be attributed to schema and labeling chaos, where the same concepts are labeled differently across separate systems. A major example of this is free text fields, as they allow for endless variations of the same input. When this inconsistency is scaled across thousands of fields and millions of records, consistent querying is made impossible without extensive cleaning. A study by the McKinsey Global Institute found that workers can spend 20 to 30 of their time on data cleansing in the absence of robust data architecture~\citep{grande2020reducing_data_costs}.

More broadly, many systems lack integration layers that connect related pieces of information. For example, PDF prints, 3D CAD models, purchase orders, and quality specifications are all critical artifacts describing the same physical object and yet they often exist in isolation, with no linking identifiers. Engineers highlight how PDF drawings do not auto-update to match approved engineering changes in CAD, or how quality specs impacting a design may not always update to reflect supplier capability changes. The CEO of Alloy Specialties reiterated how ``there’s no real way to integrate that stuff at this point… [because] everything is labeled differently, [the] data is not clean''.

\subsection{Data Representation}
\label{sec:Data_Representation}
Engineering information is represented in a variety of formats that can be poorly suited for machine consumption. Instead of critical specifications being stored in structured databases, they are often buried in requirement documents. Within PDFs, there exist quality clauses, geometric tolerances, and manufacturing process notes - all of which are essential information, and yet are inherently difficult for AI to digest. An executive at AllSpice indicates how “a lot of information in the engineering world is … locked up in PDFs,” and that ``specialized tooling [is needed] to parse data in a more standardized, structured way out of those PDFs.''  An engineer at a contract manufacturing firm reiterated this, stating ``We have lots of customers that like to find old PDFs of their bill of materials and scan them and send it to us, so [they are] not in excel files.  So you have to manually, pretty much create a bill of materials.'' These PDFs coexist with documents of all kinds such as purchase orders, 3D CAD, and quality clause documents, which each contain unique and critical information. As mentioned previously, these documents often exist in isolation and do not propagate changes - causing multiple versions to exist. Thus translation from these mixed formats into digestible and clean information is non-trivial.

Extracting data from these representations requires understanding of engineering semantics to allow for data comprehension in a robust manner. Furthermore, data parsing must be compliant with each type of file format since CAD files, for example, are software dependent. An executive at AllSpice stated the importance of ``parsing out all the schematics and PCB information, [which] enables [them] to turn it into content that even foundational models can understand''. Without a robust comprehension, downstream AIs are blind to engineer content.

On the mechanical side, there is a lack of universal standard for representing design intent. There are many competing approaches including parametric CAD, code-driven design, implicit equations, and pseudo-code representations, as mentioned by the Head of AI at Synera. Current AI systems have fundamental limitations in understanding and reasoning with 3D models. A NASA engineer highlighted how current AI tools not only struggle with assemblies and manufacturability analysis, but also with reliable comprehension and performance.

One of the most critical gaps in the mechanical workflow is automating the CAD to CAM translation. Turning geometric intent into manufacturing instructions requires a deep understanding of material properties, load cases, available tools, and machine capabilities. In the CAM workflow for machining alone, designers and operators must understand the cutting order, fixture locations, speeds/feeds, tool selection, part stability during machining, tool access, and collision avoidance. Currently this is a labor-intensive process that mandates expert manufacturing judgment and takes years of experience to do well. This also means that much of the knowledge and data is contained in undigitized formats. For example, one engineer at a contract manufacturer described relying on screenshots, handwritten notes, and memory-based workarounds to preserve information. Converting this 3D data and manufacturing know-how into machine-digestible data is still an unsolved challenge and industry is actively searching for solutions.

\subsection{Knowledge Representation}
\label{sec:Knowledge_Representation}
Most valuable manufacturing and engineering knowledge is often not codified, but rather exists in expert’s heads as intuition and experience. While standard operating procedures are documented, the intuition that experts develop over years, such as approaching difficult-to-manufacture geometries and when to bend rules, are rarely captured. 
%As highlighted by the CTO of Inkbit, ``[m]uch of the craftier knowledge about our technology is ‘in our employee’s heads’. More standardized knowledge is usually well documented/codified''. 
\begin{myquotebox}
Much of the craftier knowledge about our technology is ‘in our employees' heads.’ More standardized knowledge is usually well documented/codified.
\attributed{CTO of Inkbit.}
\end{myquotebox}
An engineer at a defense robotics company emphasized how this is exacerbated in labour intensive skilled work: ``traditional fields like fabrication are still not codified. They have a spatial and process intuition of how things will fit together and should be machined which is difficult to codify.'' Therefore, representing such advanced engineering knowledge that is derived from years of experience poses a significant data challenge. 

Even in software organizations, architectural decisions commonly exist only in the heads of senior engineers instead of the documentation. As mentioned in section~\ref{subsubsec:well-posed_tasks}, software engineers struggle with identifying the origin of variables and methods. That sort of system-level understanding and how components interact as a whole is often taught through mentorship. While documentation covers the data structures and functions within the code, it rarely covers the original intent, constraints considered, and alternatives rejected.

As mentioned previously, the sunsetting workforce demographic emphasizes the urgency of knowledge capture - especially in the aerospace and defense industries. The CEO of Alloy Specialties indicated how ``loss of institutional knowledge… [is a] significant challenge with retiring expertise and latent knowledge locked in individuals’ heads.'' Some organizations are trying to plan for this by conducting structured interviews with retiring or senior experts. This echoes a sentiment that has been long understood, tacit knowledge diffusion happens when employees leave a company, often at great loss to the company~\citep{tacit2003, DessShaw2001}. 

While challenging, this presents a real opportunity for \textbf{utilizing AI to capture and retain tacit knowledge}. As mentioned by an engineer at NASA, ``you can interview people and try to capture that knowledge, just verbally, and then you can [use] AI [to] summarize it and have it really distill… and hour conversation into… things that are truly novel''. The results can be encoded as rules in software or kept as a knowledge base for agents to access.

One challenge with captured knowledge is maintaining an up-to-date knowledge base. As emphasized by a NASA engineer, ``knowledge can get stale. [W]hat if a new adhesive comes along? How does that get  put in there?'' With the ever-adapting field of manufacturing and engineering, new materials and manufacturing methods are developed regularly. As a result, the knowledge captured from an expert five years ago may not be the current best practice, especially as industry adopts new technology. This knowledge representation poses some challenging curation questions such as identifying when captured knowledge is outdated and updating the knowledge base without re-interviewing experts.

\subsection{Data Quality}
\label{sec:Data_Quality}
Mediocre AI performance is insufficient for engineering applications. Engineering AI errors can cause part failures, safety incidents, and costly recalls and thus requires measurable performance guarantees like accuracy and repeatability, as highlighted several companies. Consistent performance of engineering AI is especially important in safety-critical decisions. To address this, some organizations employ curated knowledge graphs with verified information to ensure consistency and correctness, as further discussed in section~\ref{sec:verification_crisis}.

Part of ensuring quality data is filtering for quality training data. Just because a historical design became a real engineering solution, doesn’t mean the design is flawless. Expert designs can contain improvisations and ad-hoc decisions under pressure or without complete information. Just because a design succeeded, does not prove that the design was optimal. As reiterated by an engineer at a large engineering firm, ``do you want to train on [expert human designs]? No, you probably don’t, because the [engineer] was making a lot of stuff up, even though he’s one of our best.'' Past designs may have been developed with different tools, different computational capabilities, or different constraints, and thus using historical designs may teach the AI to mimic past compromises. One solution that organizations are proposing is to use history for inspiration and to instead use physics-based optimization with expert validation for design.

\section{Trust and Governance for Agentic AI Systems}
\label{sec: trust-governance}

Trust and governance emerged as central pillars shaping the trajectory of agentic AI adoption in engineering organizations. While technical performance matters, nearly every interviewee emphasized that no amount of model capability can substitute for trust, traceability, and robust governance structures---especially in safety‑critical, regulated, or mission‑driven domains. Across aerospace, defense, hardware, robotics, and industrial automation, organizations consistently framed AI adoption as a \textit{governance problem first}, and a \textit{technical problem second}.

\textbf{Key Takeaways:}
\begin{itemize}
\item Human-in-the-loop verification remains non-negotiable across all safety-critical and regulated domains
\item Security and data privacy requirements fundamentally shape AI architecture decisions, driving on-premise and air-gapped deployments
\item Explainability and determinism concerns persist as AI systems are compared to traditional deterministic engineering tools
\item Trust is built through institutional processes, cultural adoption, and demonstrated reliability over time
\item Organizations that align AI governance with existing engineering review structures achieve higher adoption and safer outcomes
\end{itemize}

\subsection{Human-in-the-Loop as a Non‑Negotiable Requirement} 
Our interviews indicated that AI must integrate into existing engineering processes rather than replace them, and that human-in-the-loop is a necessity. This reflects a fundamental principle: in high-stakes engineering environments, AI serves as an augmentation tool that accelerates and informs human decision-making, not as an autonomous decision-maker. As one NASA engineer explained, the institution already maintains rigorous systems to catch human mistakes before spacecraft enter orbit, and these same verification frameworks naturally extend to AI-generated outputs---but additional ones may also be necessary.

%, an executive at PTC's perspective is that ``Engineers still make the final decision, we just make explainable suggestions. It’s essential that users can see where recommendations come from.'' 

%The implementation of human-in-the-loop processes varies by company. Some treat AI as another engineering contributor whose work undergoes the same scrutiny as any human engineer's output. 

The implementation of human-in-the-loop processes was described by companies ranging from contract manufacturing to AI product development. For example, the CTO of AllSpice emphasized that for hardware design teams ``You still have humans in the loop... you're still checking with your existing process. They're not relying on the tool 100 percent.'' Meanwhile, the general manager of an electronics manufacturer also expressed openness to a human-in-the-loop model in which engineers review and validate an initial pass generated by an AI model, rather than producing all outputs manually. This approach allows teams to leverage AI's speed and analytical capabilities while maintaining the critical oversight that engineering safety demands.

\begin{myquotebox}
    Engineers still make the final decision, we just make explainable suggestions. It’s essential that users can see where recommendations come from.
    \attributed{PTC leadership team member.}
\end{myquotebox}

The validation strategies employed across organizations typically involve multiple layers of verification. Many companies implement tiered confidence systems where AI outputs are continuously monitored and compared against known ground truth cases. A senior machine learning engineer at Cyvl, which uses AI for infrastructure inspection, described their rigorous validation process: ``We're constantly retraining our models as our customers share new data. We have internal confidence in the models in addition to the evaluations on the test datasets they provide. We consistently monitor model performance to ensure top-notch outputs.'' This multi-tiered approach provides both internal quality metrics and external validation through customer feedback, creating a comprehensive trust framework.

Critically, organizations recognize that the consequences of AI errors vary significantly across different tasks. While a senior ML engineer at Cyvl noted that misclassifying the \textit{severity} of a pothole carries relatively low stakes---``It's going to require similar effort to remove it and patch it in''---other applications demand near-perfect accuracy. AllSpice's CTO articulated this clearly in the context of safety-critical hardware design: ``We really focus on the validation layer right now... These are tools that those engineers, those safety teams, can use as really powerful tools to enable them to move.'' The emphasis on validation reflects an understanding that human oversight becomes even more critical as AI is applied to higher-stakes decisions. Across organizations, AI is therefore evaluated not as an autonomous system, but as an augmented participant in long-standing engineering decision chains.

\subsection{Security, Privacy, and Deployment Architecture}
\label{sec: security-privacy-deployment}

Data security and privacy concerns fundamentally shape AI deployment architectures in engineering and manufacturing environments. The sensitivity of proprietary designs, classified information, and competitive intellectual property creates stringent requirements that go far beyond typical enterprise IT security. Nearly every organization interviewed cited data privacy as either their primary concern or a critical constraint on AI adoption, with defense contractors and regulated industries facing the most severe restrictions.

The architectural implications of these security requirements are substantial. Multiple organizations emphasized their need for on-premise deployments, air-gapped environments, and guarantees that training data remains isolated. As the head of manufacturing at an American defense technology company explained, ``Most of these algorithms basically live in the cloud or are stored in the cloud. And so there is a data problem for us, and the fact that we have to maintain our customers data. And so we are almost always looking for things that can actually be deployed on prem.'' This requirement effectively eliminates many commercial AI tools and necessitates either custom development or partnerships with vendors who can accommodate on-premise deployment models.

Defense and aerospace contractors face particularly acute security challenges due to compliance requirements such as ITAR (International Traffic in Arms Regulations) and CMMC (Cybersecurity Maturity Model Certification). For example, when asked about their company's biggest pushes for process improvement right now, an engineer at an electronics contract manufacturer noted compliance and security, and specifically complying to standards like NIST SP 800-171 or IPC-1791.\footnote{NIST SP 800-171 sets mandatory cybersecurity standards for non-federal organizations such as defense contractors to protect the confidentiality of Controlled Unclassified Information (CUI). IPC-1791, ``Trusted Electronic Designer, Fabricator, and Assembler Requirements,'' is an IPC standard and facility-level certification for ensuring security, integrity, and traceability in the electronics supply chain.}

An engineering manager at a major defense contractor described a multi-tiered security model: ``Certain levels of security require removable hard drives in a safe hand, with no external data access, or a very, very tightly controlled tunnel of data. Sometimes you have air gap systems. And air gap systems are really, really complicated to initiate and stand up.'' These requirements go beyond preventing data exfiltration---they mandate complete isolation of computing environments, making cloud-based AI services effectively unusable for many defense applications. An engineer at NASA, a \textit{civilian} space agency, also noted that IT security  and ITAR compliance are important constraints.  %``the IT security stuff is always a pain... if you're trying to use ITAR documents or something, that makes it a lot more challenging.''

Companies have responded to these constraints with various architectural approaches. A senior research and design engineer at  Autodesk described their privacy-by-design approach for their internal technology center AI research assistant: ``We're having temporary chats only... all your chat is stored on your own laptop. Autodesk has zero sense of what you're sending.'' This architecture ensures that sensitive engineering data never leaves the user's control, addressing both legal and competitive concerns. Similarly, a startup building foundation AI models has built its entire value proposition around on-device and edge deployment: ``a number of our models can run locally, so you don't need an internet connection. Your data is not getting sent up to a larger company. We work a lot with like privacy-centric enterprises like finance and healthcare, where you can't send your data out.'' This capability to run sophisticated AI models entirely on-premise or on edge devices represents a critical enabling technology for industries with strict data governance requirements.

The security concerns extend to specific use cases and data types. An employee at a defense technology startup, noted that ``we actually can't use ChatGPT for our work. We can use Copilot, but we are not allowed to use all of it because of data leakage.'' This highlights how even general-purpose coding assistants require careful evaluation and often prohibition in classified or sensitive environments. An AllSpice executive emphasized customer concerns in the hardware design space, noting that their customers want their data to remain isolated, and there is ``always reservation, [the customers] don't want [their] circuits or custom part numbers showing up [in training]. So that is very important.'' The concern is not merely theoretical---proprietary part numbers, circuit designs, and system architectures represent core competitive advantages that companies cannot risk exposing through AI training data.

Some organizations require even more granular control over data compartmentalization. An engineer at a robotic start-up noted that certain customers demand geographic and organizational isolation: ``For example, you might do [contracts] with  people in the Middle East, and [they indicate] `we need to be secure'. So in that case, we will have to provision their own cloud environment and store more data. It can't be stored with anyone else, and it has to be stored in the data center that the country is in.'' These requirements reflect not only technical security concerns but also regulatory compliance with data sovereignty laws and export control regulations.

Despite these substantial barriers, several companies demonstrated that it remains possible to leverage AI capabilities while maintaining strict security controls. AllSpice's CTO explained their approach: ``You can pass that context in in a way that's not memorized, because it is totally air gap or Sandbox like anything else. And then you can use more general design principles, like manufacture spec sheets and models and things like that.'' This architecture separates general engineering knowledge, which can come from foundation models, from company-specific proprietary context, which remains isolated. Such approaches represent the emerging best practice for balancing AI utility with security requirements.

\subsection{Explainability, Determinism, and the Engineering Mindset}
\label{sec: explainability}

A fundamental tension emerged between the probabilistic nature of many AI systems and the deterministic expectations deeply embedded in engineering culture. Traditional engineering tools and methods are valued precisely because they produce consistent, traceable, and explainable results. When engineers specify a tolerance, run a simulation, or execute a manufacturing process, they expect to understand exactly why a particular outcome occurred and to be able to reproduce that outcome reliably. This expectation of determinism creates significant cultural and technical barriers to AI adoption, particularly for systems whose decision-making processes are opaque.

An engineer at a defense robotics company articulated this concern most directly in the context of autonomous robotics: ``Software we run today is deterministic. That's how we show our work. So if something malfunctions, I can take that dataset, I can go play with it, and verify which line of code caused it to fail. If I start doing really black-box AI, it gets very difficult to figure out why.'' This ability to trace failures back to specific lines of code or specific parameter choices represents a cornerstone of engineering practice. The ``black box'' nature of many AI systems (particularly deep neural networks) fundamentally challenges this model, creating resistance even among engineers who acknowledge AI's potential benefits.

The head of manufacturing at an American defense technology company provided broader context on why this determinism matters so fundamentally: ``If you look at automation, if you look at almost anything, even compute, all of these things have been extremely deterministic. And if something is deterministic, then I can control the inputs in such a way that I can control the outputs. AI is interesting in the fact that it is not deterministic.'' This observation captures a profound shift: for the first time, engineers are being asked to trust systems that cannot guarantee identical outputs for identical inputs, that may occasionally produce inexplicable failures, and whose decision-making processes resist straightforward explanation. This represents not merely a technical challenge but a fundamental philosophical departure from engineering principles built over centuries.

\begin{myquotebox}
    For things that are a safety risk, they're in a knowledge graph so that it pulls that data, the same data the same way every time, as opposed to possibly hallucinating an answer.
    \attributed{A senior research and design engineer at Autodesk.}
\end{myquotebox}

Organizations have responded to this explainability challenge with several practical strategies. Many companies emphasized the critical importance of testing against known ground truth cases to build confidence in AI systems. An employee at a defense startup expressed this need simply: ``If there was a way to test it, to compare to ground truth, to be sure it does what it says.'' This testing-based approach allows engineers to validate AI behavior empirically, even when they cannot fully explain the underlying mechanisms. By demonstrating consistent performance across diverse test cases, AI systems can earn pragmatic trust even without perfect explainability.

%A senior research and design engineer at Autodesk described their approach to balancing AI capabilities with engineering safety requirements:``For things that are a safety risk, they're in a knowledge graph so that it pulls that data, the same data the same way every time, as opposed to possibly hallucinating an answer.''
Some organizations have gone further by architecting their AI systems to maximize explainability wherever possible. As explained in the quote above by an engineer at Autodesk, hybrid approaches can leverage the non-deterministic capabilities of large language models for lower-stakes applications while using deterministic knowledge retrieval for safety-critical information. Such approaches acknowledge that complete determinism may not always be achievable while ensuring that the highest-stakes decisions remain traceable and reproducible.

The explainability challenge also manifests in documentation and compliance requirements. Engineering work must be documented, auditable, and defensible to regulators, insurers, and in legal proceedings. As several interviewees noted, traceability requirements drive several practical constraints: AI outputs must be logged, tool actions must be replayable, workflows must be attributable, and AI-generated changes must provide diffs showing exactly what changed rather than simply presenting final results. For regulated industries, this is not optional---it is fundamental to their ability to operate and maintain certification.

Several companies emphasized that building trust in AI systems requires patient demonstration of reliability over time. A manufacturing engineer at Saunders Electronics described a pragmatic openness to AI, noting that tools are adopted quickly when they are reliable, save time, and integrate seamlessly into existing workflows. Overcoming skepticism requires AI systems to demonstrate consistent, explainable behavior across hundreds or thousands of use cases, gradually building the empirical foundation for trust.

\subsection{Aligning with and Adding to Established Engineering Review Processes}

Organizations that successfully integrate AI into their workflows do so by aligning AI governance with established engineering review cultures rather than attempting to replace or circumvent existing processes. This alignment reflects a pragmatic recognition that engineering organizations have developed their review structures over decades to ensure safety, quality, and regulatory compliance. Rather than viewing these structures as obstacles to innovation, leading organizations treat them as frameworks for responsible AI adoption.

A NASA engineer articulated this principle clearly in the context of spacecraft development: ``We already have these processes, PDR (Preliminary Design Review), CDR (Critical Design Review), TRR (Test Readiness Review), nothing about AI suggests abandoning them,'' however, ``we might need to augment our review processes with both AI-automated review and more frequent human-in-the-loop review as complex systems are developed ever-faster in the `Agentic Engineering' future.'' We can see that these existing milestone reviews represent critical checkpoints where engineering decisions undergo rigorous scrutiny by cross-functional teams including safety engineers, quality assurance specialists, and subject matter experts. By positioning AI-generated outputs as inputs to these existing reviews rather than as replacements for human decision-making, engineering and manufacturing firms can maintain institutional safety culture while leveraging AI's analytical capabilities. 

This approach extends across organizations and engineering disciplines. An engineering manager at a major defense contractor described how they extend their standard review checkpoints to AI-generated designs: ``We use the same checkpoints for AI decisions as we do for human decisions. If it passes the logic reviews, it gets approved.'' This framework treats AI as another contributor whose work undergoes the same validation processes that apply to human engineers. This approach is simple to deploy: rather than creating entirely new governance structures for AI, organizations leverage decades of accumulated wisdom about engineering validation. That said, as AI increases the volume and speed of design iteration, organizations may need to introduce dedicated validation checkpoints for AI-generated outputs to ensure quality scales alongside quantity.

The integration of AI into engineering review processes also addresses a practical concern: how to maintain engineering judgment and expertise even as AI tools become more capable. An energy equipment manufacturing and services company emphasized their focus on ``improving shop floor efficiency and operator experience before considering fully autonomous applications.'' This staged approach allows organizations to validate AI capabilities in lower-stakes applications before gradually expanding to higher-risk use cases. It also ensures that human engineers maintain active engagement with engineering problems rather than becoming passive supervisors of automated systems.

Several organizations noted that this alignment with existing processes actually accelerates adoption by reducing organizational resistance. When AI appears as an extension of engineering rigor rather than a disruption to it, engineers become more willing to experiment with and validate AI tools. A senior research and design engineer at Autodesk described their strategy of identifying pain points within existing workflows and deploying AI to address specific friction points: ``We can ask Tech Center staff and research engineers what pipelines or what workflows they find annoying to do, and find ways to do them with the AI assistant.'' This targeted approach generates quick wins that demonstrate value while respecting the overall engineering process architecture.

The review culture also provides natural checkpoints for monitoring AI system performance over time. As organizations deploy AI tools more broadly, they accumulate data on AI accuracy, failure modes, and areas where human oversight remains essential. This empirical feedback allows for continuous refinement of governance policies based on observed performance rather than theoretical risk assessments. Several companies emphasized the importance of this learning process, noting that their governance frameworks have evolved substantially as they gained operational experience with AI systems.

\subsection{Building Trust Through Organizational Culture and Communities of Practice}
\label{sec: building-trust-orgs}

While formal governance policies and technical safeguards provide essential foundations for AI adoption, multiple organizations emphasized that trust ultimately grows through cultural mechanisms: hands-on experience, peer learning, visible leadership support, and communities of practice that share knowledge and build collective expertise. Technical capabilities and policy frameworks matter, but the human dimension of trust-building often determines whether AI adoption succeeds or stalls.

A NASA engineer reported significant grassroots engagement with their internal AI platform, ChatGSFC: ``We have a community of practice meeting, we have over 6,000 people signed up for our ChatGSFC, we have AI champions ... but some people that are down in the trenches, it's hard to reach them.'' This highlights a common pattern across organizations: while early adopters and AI enthusiasts quickly embrace new tools, broader organizational adoption requires sustained effort to reach skeptics and those focused on day-to-day execution rather than exploration of new technologies. NASA's approach of designating ``AI champions'' throughout the organization represents one strategy for bridging this gap, creating trusted peers who can provide hands-on guidance and vouch for AI capabilities based on direct experience.

Organizations identified distinct adoption personas that shape how trust develops across their workforce. One leader described three categories: (1) enthusiasts who overestimate AI’s capabilities and expect it to work universally, (2) skeptics who abandon it after an initial failure, and (3) pragmatic users who recognize its limitations and derive value by applying it judiciously. The challenge for organizations lies in moving people from the extremes toward this middle ground of informed, pragmatic use. This requires not just access to tools but active support in developing appropriate mental models of AI capabilities and limitations.

The importance of leadership engagement emerged as a consistent theme. As an engineer at NASA noted, ``You have to get management buy-in from the very top level. Companies where the CEO is using it are much more likely to adopt it than companies where they're not, because they can just see it as a distraction or a risk if they don't really, at a gut level, understand the value.'' This observation aligns with broader change management research: transformative technologies require visible leadership support to overcome organizational inertia and risk aversion. When leadership actively uses AI tools and discusses their value, it signals organizational commitment and provides ``permission'' for others to invest time in learning new approaches.

However, leadership support alone does not guarantee successful adoption. Organizations emphasized the critical importance of peer-to-peer learning and organic knowledge sharing. Reaching frontline engineers requires sustained effort, as many have yet to experience the kind of personal ``aha moment” that builds trust in AI. Organizations can encourage these moments by providing accessible tools, sharing success stories, and giving employees time to experiment. Some have introduced structured training sessions, such as focused AI practice days with hands-on, over-the-shoulder coaching, which help employees quickly develop practical skill and confidence.

Trust-building also requires an honest acknowledgment of limitations and failures. Several organizations emphasized the danger of overhyping AI capabilities, which can lead to disillusionment when tools fail to meet inflated expectations. One engineer at a robotic start-up captured this sentiment: ``I'm pretty skeptical. Personally, I don't use copilot, I don't use structure here, just because I feel like I end up having to rewrite it.'' Such skepticism often stems from early negative experiences with AI tools. Organizations that succeed in building trust acknowledge these limitations openly and help users develop realistic expectations about when AI adds value and when traditional methods remain superior.

The cultural dimension of trust-building intersects closely with technical governance. An engineer at Cyvl described a customer who initially hesitated to adopt an AI tool until they were shown that they could download and inspect the exact data the system used and validate it within their own software. That level of transparency quickly shifted the customer’s perception and accelerated adoption. This example illustrates how features that expose data sources, enable output verification, and clarify system behavior can meaningfully strengthen cultural trust by giving skeptics concrete ways to assess AI performance for themselves.

Ultimately, organizations recognized that governance operates at two levels simultaneously: as formal policy that defines acceptable use, security requirements, and validation processes, and as social practice that shapes daily decisions about whether and how to use AI tools. The most successful organizations attend to both dimensions, recognizing that technical safeguards without cultural support lead to compliance without engagement, while cultural enthusiasm without governance structures creates unacceptable risk. As one interviewee summarized, governance is therefore both a policy layer and a social adoption layer, requiring sustained attention to both technical and human factors.

% \textbf{Section Summary:}

% Trust and governance in AI-enabled engineering environments are not merely technical problems to be solved through better algorithms or stricter access controls. Rather, they represent multifaceted challenges requiring alignment of technical architecture, organizational policy, regulatory compliance, and cultural change. The organizations achieving success in AI adoption share several common characteristics: they maintain human-in-the-loop oversight for all safety-critical applications, they architect systems to meet stringent security and privacy requirements, they prioritize explainability and determinism even at the cost of some AI capabilities, they align AI governance with established engineering review processes rather than attempting to replace them, and they invest in cultural trust-building through communities of practice, peer learning, and leadership engagement.

% These insights reveal that the ``governance problem'' of AI in engineering is not a temporary barrier to be overcome en route to full automation, but rather a permanent feature of how organizations responsibly deploy powerful but imperfect tools in high-stakes environments. Organizations that recognize this reality and build governance frameworks accordingly---frameworks that combine technical safeguards with cultural practices, that respect engineering traditions while enabling innovation, and that maintain human judgment as central to engineering decisions---position themselves to realize AI's substantial benefits while avoiding its risks.

\section{Barriers to the Adoption of Agentic Systems}
\label{sec: Barriers}

Interviewees highlighted a consistent set of barriers preventing AI and agentic systems from reaching their full potential in engineering workflows. These barriers span technical, cultural, organizational, and infrastructural challenges, revealing that the path to widespread AI adoption in engineering is not merely a question of improving model performance but rather requires coordinated advances across multiple dimensions simultaneously. The interviews revealed a few central themes that, when addressed, could unlock substantially greater AI utility in engineering and manufacturing contexts.

\textbf{Key Takeaways:}
\begin{itemize}
\item Engineering-grade reliability and verifiability remain the most critical unmet technical requirements, with organizations demanding testable, repeatable, and explainable AI behavior for safety-critical applications
\item A fundamental AI literacy gap persists across the engineering workforce, with few engineers understanding how to build, configure, or critically evaluate AI systems beyond basic usage
\item Current AI models lack robust spatial reasoning and multi-physics understanding essential for engineering tasks, limiting their applicability to integrated design problems
\item Legacy engineering tools were not architected for AI integration, lacking APIs, programmatic interfaces, and structured data access that modern AI systems require
\item Organizational culture varies dramatically from risk-averse conservatism to move-fast startup mentality, with both extremes presenting distinct adoption challenges
%\item Key breakthroughs needed include standardized agent interfaces, verification frameworks, spatial reasoning models, data transformation pipelines, and cultural adoption mechanisms
\end{itemize}

\subsection{Verification and Reliability Gap}
\label{sec:verification_crisis}

A widely cited barrier across all interviews was the absence of engineering-grade standards for AI reliability, verifiability, and repeatability. While AI systems have demonstrated impressive capabilities in many domains, engineering organizations require a level of provable, testable behavior that current AI technologies often cannot provide. This gap represents not only a technical limitation but a fundamental mismatch between how engineering has historically validated systems and how modern AI operates.

Traditional engineering validation relies on deterministic systems whose behavior can be precisely predicted and repeated. As a result, engineering fields often struggle with the uncertainty inherent in modern AI.
%However, as a leader at PTC points out, ``engineers want deterministic behavior. LLMs are probabilistic by nature, so balancing those two expectations is one of the hardest technical challenges.''
\begin{myquotebox}
Engineers want deterministic behavior. LLMs are probabilistic by nature, so balancing those two expectations is one of the hardest technical challenges.
\attributed{PTC leadership team member.}
\end{myquotebox}

When an engineer specifies a tolerance, runs a structural analysis, or programs a CNC machine, they expect identical inputs to produce identical outputs every time. This determinism enables comprehensive testing regimes, formal verification methods, and clear assignment of responsibility when failures occur. AI systems, particularly those based on large language models, fundamentally challenge this paradigm through their probabilistic nature and occasional unpredictable behavior.

A senior research and design engineer at the Autodesk Technology Center, articulated the enterprise constraint succinctly: ``You have to benchmark its accuracy and know how accurate it's going to be.'' This requirement extends beyond merely knowing average performance---organizations need to understand failure modes, quantify uncertainty, establish confidence intervals, and predict under what conditions AI systems might produce incorrect or unsafe outputs. Without this level of characterization, AI remains confined to applications where errors carry minimal consequences.

The stakes become dramatically higher when AI systems control physical processes or make safety-critical decisions. An engineer from a defense robotics company emphasized the critical importance of root cause analysis: ``If something malfunctions, we need to find the exact line of code that caused it. With black-box AI that's much harder.'' This statement captures a fundamental tension: engineering cultures built around deterministic debugging and clear causation must now grapple with systems whose decision-making processes resist straightforward explanation. When an autonomous system fails, regulators, insurers, and customers demand detailed explanations of what went wrong and proof that it won't recur; requirements that current AI architectures does not fully satisfy.

An engineer at a defense technology company, made the same point from a different angle: ``We need a way to test it, to compare it against ground truth.'' This reflects a broader pattern where organizations want to validate AI systems the same way they validate any engineering system: through rigorous testing against known correct answers, stress testing at boundary conditions, and formal verification of critical properties. The challenge lies in defining appropriate test suites for AI systems, particularly when those systems are meant to handle novel situations that couldn't be enumerated in advance.

The verification challenge manifests differently across application domains. For code generation and documentation tasks, where errors are often easily detected and corrected, looser verification standards may suffice. However, for structural design, safety analysis, or manufacturing process control, the consequences of AI errors demand substantially higher reliability thresholds. Organizations consistently reported that they cannot deploy AI for mission-critical applications until verification methods mature to match the standards already established for traditional engineering tools.

Several organizations described their attempts to develop internal verification frameworks. These typically involve combinations of redundant checking, human validation of critical outputs, extensive logging of AI decision-making processes, and mechanisms to replay and debug AI actions. However, these ad hoc solutions represent significant engineering overhead and don't fully address the underlying explainability challenges inherent in modern AI architectures. Until the research community and tool vendors develop standardized, widely accepted verification frameworks specifically designed for AI in engineering contexts, reliability concerns will continue to limit AI adoption for high-stakes applications.

\subsection{AI Literacy Gap and Workforce Development Challenges}
\label{sec:ai-literacy}

Beyond technical limitations of AI systems themselves, organizations identified a profound literacy gap across their engineering workforce as a major barrier to adoption. While a growing number of engineers have experimented with AI tools, very few possess the deeper understanding necessary to critically evaluate AI capabilities, recognize when AI is appropriate or inappropriate for particular tasks, troubleshoot AI failures, or configure AI systems for specialized engineering applications. This knowledge gap manifests at multiple levels and significantly constrains how effectively organizations can deploy AI technologies.

\begin{myquotebox}
Many know how to use AI, very few know how to build or reconfigure it.
\attributed{An executive at a midsized engineering firm.}
\end{myquotebox}

%An  captured this challenge with stark clarity: ``Many know how to use AI, very few know how to build or reconfigure it.'' This distinction proves critical in practice. 
Surface-level AI usage---asking questions of ChatGPT, using code completion tools, generating simple text---requires minimal expertise and has become relatively widespread. However, effectively integrating AI into engineering workflows demands substantially deeper understanding: knowledge of model architectures and their limitations, ability to evaluate output quality and detect hallucinations, understanding of when to trust AI suggestions versus when human judgment should override them, and capacity to configure and fine-tune systems for specific engineering contexts.

The literacy gap creates several downstream challenges. Organizations struggle to assess vendor claims about AI capabilities, making procurement decisions difficult. They cannot effectively evaluate whether poor AI performance stems from fundamental model limitations or from improper configuration and usage. They have difficulty identifying which engineering tasks are well-suited for AI automation versus which should remain primarily human-driven. Most critically, they cannot train their workforce at scale because they lack sufficient internal expertise to develop comprehensive training programs.

Even in highly resourced, technologically sophisticated organizations, building AI literacy across the full workforce is a gradual and uneven process. NASA, as described in Section~\ref{sec: building-trust-orgs}, has invested significantly in internal AI infrastructure, including an internal language model (ChatGSFC), and it has created a community of over 6,000s users, but it is still difficult to reach those employees ``down in the trenches.'' Efforts like these build the institutional scaffolding for AI adoption, even as they highlight that meaningful integration ultimately happens when individuals see clear value in their own workflows.

The adoption dynamics described above resonate broadly across organizations: ``You have people who think it solves everything, people who think it's useless, and the middle ground---the one we want---who use it while understanding its limitations.'' This tripartite division reflects different mental models of AI capabilities. Overly enthusiastic early adopters may trust AI inappropriately, potentially introducing errors into critical workflows. Skeptics who dismiss AI entirely after one negative experience deny their organizations potential productivity gains. The desired middle ground---informed, critical use that leverages AI strengths while compensating for weaknesses---requires precisely the kind of nuanced understanding that remains rare across engineering workforces.

Several factors contribute to the persistence of this literacy gap. Engineering education has not yet broadly integrated AI literacy into core curricula, meaning recent graduates often lack relevant skills despite their technical sophistication in traditional engineering disciplines. Professional development programs struggle to keep pace with rapidly evolving AI capabilities, and many engineers face time constraints that prevent deep engagement with new technologies. The shortage of instructors with combined AI and engineering expertise further constrains training capacity.

Organizations attempting to address these challenges through internal training programs face difficult pedagogical questions. Should training focus on high-level concepts and appropriate use cases, or dive into technical details of model architectures? How can training accommodate engineers with vastly different baseline technical sophistication and learning preferences? What hands-on practice opportunities allow engineers to develop intuition about AI capabilities and limitations? These questions lack settled answers, and most organizations are still experimenting with different approaches.

The observation from a software engineer from LangChain about tool development reflects awareness of this challenge: ``Our tech stack primarily caters to developers but we are trying to expand to low-code, no-code tools that we build agents with. We are looking to make the jump so that non-developers can also build.'' This represents one promising approach: creating abstractions that allow engineers to leverage AI capabilities without requiring deep machine learning expertise. However, such abstractions inevitably involve tradeoffs---they may limit customization, hide important details about system behavior, or create their own learning curves. The optimal balance between accessibility and power remains an open question.

The literacy gap also intersects with trust and adoption dynamics discussed earlier. Engineers who lack deep AI understanding may struggle to develop appropriate trust calibration---neither blind faith nor blanket skepticism, but rather nuanced judgment about when AI outputs merit confidence. This calibration typically develops through extensive hands-on experience, which many organizations struggle to provide systematically across large engineering workforces.

\subsection{Spatial Reasoning and Multi-Domain Understanding Challenges}
\label{sec:Barrier_spatial}
%\todo[inline]{Faez: Strengthen the evidence presentation. include a table mapping each theme to how many interviewees mentioned it and which stakeholder groups emphasized it.}

Current foundation models demonstrate impressive language, code, and certain visual capabilities but fall consistently short at the spatial reasoning, multi-physics integration, and systems-level thinking that characterize engineering work. This is not merely an incremental limitation but potentially a fundamental architectural challenge.

%/\subsubsection{Spatial Reasoning}
\begin{myquotebox}
Design understanding plus spatial understanding is super important. Nothing exists alone; it's always part of a bigger system. So we always need spatial understanding.
\attributed{Head of AI at Synera.}
\end{myquotebox}

%As the Head of AI at Synera articulated, spatial understanding is critical. He added that ``nothing exists alone; it's always part of a bigger system. So we always need spatial understanding.'' This observation captures a reality that every mechanical engineer intuitively understands but that current AI systems struggle to grasp: 
Engineering components exist in three-dimensional space, interact with surrounding structures, must accommodate manufacturing constraints, and participate in larger assemblies with complex geometric and functional relationships. Throughout the engineering design process, a design moves through many modalities including hand drawn sketches, textual descriptions, parametric 3D models, meshes for use in simulations, physical prototypes, sub-assemblies, and finally a manufactured product~\citep{ulrich2020product}. An AI system that can describe a bracket in text but cannot reason about how that bracket interfaces with surrounding components, whether it can be accessed for assembly, or how it distributes loads within a larger structure provides limited utility for actual engineering work.

The challenge extends beyond geometric understanding to encompass multi-physics reasoning. Real engineering problems typically involve coupled phenomena across multiple domains simultaneously, as shown in other exploratory works~\citep{ferdous2024automation, Picard2025}. 

An engineering manager at a major defense contractor described requirements analysis for complex systems that must consider thermal management, electrical power distribution, mechanical loads, software control systems, and manufacturing constraints all at once. As one engineering manager at a major defense contractor explained, understanding ``what happens when the vehicle gets too hot'' requires reasoning across software throttling decisions, electrical power routing, thermal dissipation through mechanical structures, and the physical constraints of sealed enclosures---a level of integrated systems thinking that current AI models cannot replicate.

The spatial reasoning gap manifests in several specific ways. Current vision-language models can describe images but struggle with precise spatial relationships, distance estimation, or reasoning about hidden geometry. They cannot reliably answer questions like ``will this part clear the adjacent component during assembly?'' or ``what is the shortest collision-free path for this robot arm?'' 3D CAD models, which represent the native language of mechanical engineering, remain largely opaque to current foundation models. While some specialized systems can process point clouds or mesh representations, they typically lack the semantic understanding that allows human engineers to recognize design intent, identify functional relationships, and reason about design alternatives.

Several organizations mentioned the data scarcity challenge that compounds spatial reasoning difficulties. The Head of AI at Synera noted that ``the time taken to generate a design is larger than a time taken to generate a code snippet. Even if OpenAI had to build a completely new evaluation benchmark, they could probably do it faster than [the time] it would take thousands of mechanical engineers to design parts because the time to create one sample is much larger for design.'' This observation highlights a fundamental challenge: while software training data exists in vast quantities and can be synthetically generated at scale, high-quality engineering design data remains scarce, proprietary, and expensive to create. The massive datasets that enabled breakthrough performance in language and even code generation simply don't exist for mechanical design, and generating them would require extraordinary effort.

This data scarcity also reflects the diversity and specialization within engineering domains. A software foundation model can be trained on code from millions of repositories covering thousands of programming tasks, achieving broad generalization. But mechanical engineering encompasses vastly different design paradigms---from aerospace structures optimized for weight under stress constraints, to injection-molded consumer products optimized for manufacturability and cost, to precision optical instruments with nanometer-scale tolerances. Each domain involves different materials, manufacturing processes, analysis methods, and design principles. Building AI systems with genuine engineering competence across this diversity would require not just more data but fundamentally better approaches to transfer learning and domain adaptation.

Some organizations described promising but still limited progress on specific subproblems. The CEO of from Alloy Specialties noted the challenges: ``You can't just deploy a Spotify or Facebook nearest neighbor sort of similarity tool, because you're going to get a part that's this big and a part that's \textit{this big}, and [the algorithms are] going to be like 'these two parts are the same' - and they're clearly not. How you make this part and how you make that part [are] two very, very different things.'' This example illustrates how engineering similarity involves not just geometric shape but scale, manufacturing history, material properties, tolerance requirements, and functional intent---a much richer semantic space than typical similarity metrics capture.

The path forward likely lies in hybrid architectures that combine foundation model capabilities with specialized engineering reasoning systems. Cutting-edge research is developing AI that can work with parametric CAD~\citep{cadcoder2025doris, alam2025gencad}, generate 3D meshes from sketches~\citep{Edwards_Man_Ahmed_2024} or other 2D images~\cite{lai2025hunyuan3d25highfidelity3d, jun2023shape}, and embed physics-informed reasoning for domains such as fluid dynamics~\citep{Cai2021PINNs}. However, these advances largely remain research prototypes, with limited integration into the validated, security-constrained, and compliance-driven workflows of industrial engineering practice. Some organizations experiment with architectures that use language models for high-level reasoning while delegating geometric queries to traditional CAD kernels, or that augment LLMs with structured engineering knowledge graphs. Yet these approaches remain exploratory, and fundamental questions about system reliability, verification, and architectural design remain unresolved.

\subsection{Legacy Tools Integration Problems}
\label{sec: legacy-tools}

A near-universal barrier highlighted across interviews was the profound mismatch between how modern AI systems expect to interact with software and how legacy engineering tools were actually architected. Engineering software ecosystems evolved over decades with assumptions about human interaction, file-based workflows, and GUI-driven operation that fundamentally conflict with the programmatic access, real-time data exchange, and structured interfaces that AI agents require. This legacy integration challenge affects not just occasional friction points but represents a systemic barrier to AI adoption across the engineering toolchain.

One research engineer at a large enterprise captured the inertia that many organizations face: ``We've been using Pro Engineer since the '90s and changing tools is incredibly difficult.'' This statement reflects more than simple resistance to change. Engineering tools become deeply embedded in organizational processes, with customizations, extensions, legacy data, skilled personnel, and regulatory certifications all tied to specific software platforms. The switching costs---in terms of capital investment, workflow disruption, retraining, and risk---are immense. Organizations cannot simply abandon decades of accumulated tooling investment, yet their existing tools often lack the interfaces necessary for meaningful AI integration.

Synera's Head of AI framed the gap in terms of mismatched evolutionary timescales: ``The speed at which LLMs are growing and the speed at which engineering tool ecosystems are growing aren't matching. There is a huge delta.'' This observation highlights a fundamental problem: AI capabilities are advancing on Moore's Law-like exponential curves, while industrial software evolves incrementally through multi-year development cycles driven by conservative customer requirements. Tool vendors must balance innovation against backward compatibility, regulatory compliance, and reliability expectations that make rapid architectural changes impractical. The result is a growing capability gap between what AI systems could theoretically do and what legacy tools actually expose.

The specific technical barriers manifest in predictable patterns. Most engineering software lacks REST APIs or any programmatic interface designed for external automation. Tools were built assuming human operators working through graphical interfaces, with critical functionality accessible only via mouse clicks, menu navigation, and visual feedback. When automation capabilities exist at all, they typically take the form of macro scripting languages or batch modes designed for simple repetitive tasks rather than the sophisticated tool orchestration that AI agents require. Many tools cannot operate headlessly without full GUI initialization, making them unsuitable for cloud deployment or background automation. Data exchange between tools typically occurs through file exports rather than structured data streams, introducing latency and serialization overhead that limit real-time AI interaction.

The lack of structured logs and audit trails further complicates AI integration. When engineers interact with tools manually, implicit knowledge guides their actions---they understand why they clicked a particular button or chose a specific parameter. AI systems attempting to automate these workflows need explicit traces of decision logic, parameter rationale, and success criteria. Without such structured information, AI cannot learn from human demonstrations, cannot explain its own actions in terms humans understand, and cannot support the traceability requirements that governance demands.

Several organizations described workarounds they developed to bridge these gaps. A senior research and design engineer at Autodesk explained their approach: ``We want everything to happen in the same environment. So we built plugins with a bunch of companies that are providing that process optimization.'' This plugin-based approach allows AI capabilities to be embedded directly within existing tools rather than requiring wholesale replacement. However, developing plugins for dozens of disparate tools, each with different extension mechanisms and capabilities, represents substantial engineering effort. Moreover, plugin-based integration typically cannot access the deepest tool functionality or modify core algorithms, limiting what AI can ultimately accomplish.

Some forward-thinking tool vendors are beginning to address these limitations. Synera's Head of AI noted that ``key enablers include action logging, open APIs, and clear ontologies for tool-to-agent translation.'' These represent exactly the architectural features that AI-friendly engineering tools would provide: comprehensive APIs to share data programmatically, detailed logging of all operations for training and debugging, and well-defined semantic schemas that allow AI systems to understand tool capabilities and data structures. The Model Context Protocol (MCP), mentioned by several interviewees, represents one emerging standard attempting to provide common interfaces for AI-tool integration. However, widespread MCP adoption remains aspirational, and even tools that do provide APIs often expose incomplete or inconsistent functionality.

The integration challenge also extends to data transformation. Engineering knowledge exists in myriad formats: CAD models in proprietary formats, PDF drawings, spreadsheet calculations, tribal knowledge in engineers' heads, photographs of physical prototypes, and handwritten notes. The CEO of Alloy Specialties emphasized the difficulty of making this diverse information accessible to AI: ``Every single prime has a different series of quality clauses. They have a different series of requirements from business unit to business unit. Being able to ingest and then input into an ERP system all of the engineering data that that's not 3D model data, plating specs, quality specs, inspection requirements, everything else that's on the left hand side of that print.'' Creating data transformation pipelines that can reliably extract, structure, and contextualize this information at scale remains a significant technical challenge.

Organizations face difficult strategic decisions about how to address tool integration barriers. Should they invest in extensive custom integration with existing tools, accepting limited functionality? Should they pressure vendors to modernize their platforms, accepting slow progress given vendor priorities? Should they consider wholesale migration to modern, API-first tools, accepting massive switching costs and risk? Or should they implement AI in parallel workflows that don't directly integrate with existing tools, accepting additional manual data transfer overhead? Each organization answers these questions differently based on their specific constraints, and no clear consensus has emerged about optimal approaches.

\subsection{Organizational Culture Spectrum: A Balance of Innovation and Risk}
\label{sec: org-culture-spectrum}

Organizations' cultural attitudes toward new technology and risk create powerful forces that either accelerate or impede AI adoption, often independent of technical readiness or economic justification. The interviews revealed a spectrum of organizational cultures, from deeply conservative institutions where change requires extensive validation and consensus, to fast-moving startups where the greater risk lies in moving too slowly to capitalize on emerging capabilities. %Our interviews suggest that neither extreme proves optimal for sustainable AI adoption, and organizations struggle to find appropriate middle ground.
\begin{myquotebox}
    We can either stay deterministic and be a 1997 company, or we can figure out how to use this stuff and be a 2030 company.
    \attributed{An engineer at a defense robotics company.}
\end{myquotebox}

%An engineer at a defense robotics company articulated the tension many organizations feel: ``We can either stay deterministic and be a 1997 company, or we can figure out how to use this stuff and be a 2030 company.'' This framing captures the high stakes that some leaders perceive: AI adoption is not merely about incremental efficiency gains but about fundamental competitive viability. Organizations that fail to modernize may find themselves unable to compete with more agile competitors who leverage AI for faster development cycles, lower costs, or superior products. This fear of obsolescence can drive aggressive AI adoption even when technical maturity might counsel caution.

%However, this pressure to move fast conflicts with legitimate concerns about reliability, safety, and regulatory compliance that characterize engineering-heavy industries.
An engineer at a large engineering enterprise described the cultural conservatism that develops in organizations with long experience of technology hype cycles: ``We're always getting new technologies thrown at us, it makes people skeptical.'' This skepticism doesn't necessarily reflect resistance to innovation but rather hard-earned wisdom about distinguishing genuine transformative technologies from temporary fads. Organizations in aerospace, defense, and other regulated sectors have seen countless ``revolutionary'' technologies that promised transformation but delivered modest improvements at best. Their caution reflects rational risk management given the consequences of premature adoption of immature technologies.

The cultural challenge manifests differently across organizational types. Large, established enterprises typically face layers of approval processes, compliance requirements, and established best practices that slow any change---AI or otherwise. Multiple business units may need to coordinate, security teams must review new tools, procurement processes take months, and pilot programs require extensive validation before broader rollout. While this deliberate approach reduces risk, it also means these organizations often lag years behind cutting edge capabilities. Some interviewees from large companies expressed frustration about moving slowly even when technical teams recognized clear AI opportunities.

Small, venture-backed startups face opposite pressures. One such startup captured this challenge: ``The biggest challenge is we move so fast we don't have time to fit AI smoothly into the workflow.'' In fast-moving environments, pausing to properly integrate new capabilities, develop appropriate governance frameworks, or comprehensively train teams feels like an unaffordable luxury. This sentiment is echoed in prior research showing that startups with markers of high-growth entrepreneurship (such as venture capital funding, high initial capitalization, and
reliance on formal intellectual property) had the highest AI usage~\citep{McElheran2024AIAdoptionAmerica}.

%Teams ship products with AI components that have not been thoroughly validated, adopt new tools without proper security review, and accept technical debt with plans to refine ``later'' that often never materialize. While this velocity enables rapid iteration and learning, it also creates sustainability risks and potential quality or security incidents.

%Neither extreme proves sustainable. Organizations that move too conservatively find their talented engineers frustrated by bureaucracy and increasingly drawn to more dynamic competitors. They accumulate technical debt of a different sort: legacy processes and tools that become increasingly difficult to maintain as the broader ecosystem evolves. Meanwhile, organizations that move recklessly without appropriate governance eventually face safety incidents, security breaches, or quality failures that force painful course corrections and erode customer trust.

Some organizations have developed cultural mechanisms to balance innovation and stability. An engineer at an energy equipment manufacturing and services company described their philosophy: ``There's an encouragement to use the tools, it's not like Thou shalt integrate AI, or thou shalt not. It's always like, it's a tool.'' Treating AI as another tool in the engineering toolkit rather than as either mandate or prohibition creates space for experimentation while maintaining individual judgment about appropriate use. Engineers feel empowered to try AI for suitable applications without pressure to apply it universally regardless of fit.

One engineer at an energy equipment manufacturing and services company  also emphasized the importance of pragmatic validation: ``You don't want to shoot a fly with a cannon, so if I don't have to build a neural network to do one little test, I'm not gonna do it.'' This principle of using the simplest effective solution rather than the most sophisticated helps organizations avoid both over-engineering and missing genuine opportunities. It requires engineers to understand both AI capabilities and traditional methods well enough to make informed choices about which approach fits each specific problem.

Several interviewees emphasized that successful cultural evolution requires patience and sustained effort rather than mandates or quick fixes. One NASA engineer's observation that ``everybody needs to have their own aha moment, but a lot of people have not had that'' reflects recognition that individual conversion experiences, where someone directly sees AI solve their problem, often prove more powerful than top-down directives. Creating conditions where such moments can occur---providing easy tool access, sharing success stories, supporting experimentation, accepting some failures---represents a long-term cultural investment rather than a quick organizational fix.

The cultural challenge also intersects with workforce development issues discussed earlier. Organizations with conservative cultures may struggle to attract and retain AI-savvy engineers who want to work with cutting-edge technologies. Conversely, organizations that move too fast may burn out employees with constant churn and inadequate support for learning new systems. Finding cultural balance proves essential not just for effective AI adoption but for organizational sustainability.

\section{Key Breakthroughs That Could Transform AI and Agents Utility in Engineering}
\label{sec:Breakthroughs_benefit}

While the barriers outlined above may seem daunting, interviews also surfaced a clear consensus about the key breakthroughs that could dramatically expand AI utility in engineering contexts. These represent not merely incremental improvements but potential step-function changes in capabilities that would unlock substantial new applications. Importantly, several of these breakthroughs span both technical and socio-technical domains, reflecting the reality that AI adoption barriers are not purely technical.

\paragraph{1. Standardized Agent Interfaces and Tool Schemas}

A frequently mentioned technical need was for standardized, widely adopted interfaces that allow AI agents to interact with engineering tools programmatically. Rather than each AI system and each tool developing custom integrations, the field needs agreed-upon protocols analogous to how HTTP enabled web applications or SQL standardized database access. The Model Context Protocol (MCP) represents one promising effort in this direction, but widespread adoption remains uncertain. Such standards would need to address not just API conventions but also semantic schemas that define how to describe tool capabilities, data types, operations, and error conditions in ways that both humans and AI systems can understand.

% The Head of AI at Synera emphasized that ``key enablers include action logging, open APIs, and clear ontologies for tool-to-agent translation.'' This combination proves essential: APIs provide the mechanism for agents to invoke tool functionality, action logging creates the audit trails necessary for debugging and compliance, and ontologies enable semantic understanding that allows agents to reason about what tools can do and compose them appropriately. Without all three elements, agent-tool integration remains brittle and application-specific.

\paragraph{2. Robust Verification Frameworks Specifically for AI in Engineering}

As discussed extensively earlier, the absence of engineering-grade verification methods represents perhaps the single greatest barrier to AI adoption in safety-critical applications. The breakthrough needed extends beyond simply testing AI systems more thoroughly to developing fundamentally new verification paradigms suited to probabilistic, non-deterministic systems operating in complex environments.

Such frameworks would likely combine multiple approaches: simulation-based validation where AI systems are tested in high-fidelity virtual environments across thousands of scenarios; formal methods adapted to reason about probabilistic behavior and quantify uncertainty; runtime monitoring that detects anomalous AI behavior during operation; and deterministic replay mechanisms that allow exact reconstruction of AI decisions for debugging. The frameworks must also address how to decompose complex AI systems into verifiable components, how to establish safety envelopes within which AI can operate autonomously, and how to define appropriate human oversight requirements for different risk levels.

Several organizations described developing internal verification approaches, but the lack of industry-wide standards means each organization essentially starts from scratch. Collaborative development of verification frameworks, potentially through standards bodies or industry consortia, could dramatically accelerate progress.

\paragraph{3. AI Models with Native Spatial and Multi-Physics Reasoning}

Current foundation models' weakness in spatial reasoning and multi-domain integration limits their engineering utility despite impressive language capabilities. The breakthrough needed goes beyond simply adding more geometry data to training sets---it likely requires fundamental architectural innovations that allow models to represent and reason about 3D space, physical constraints, and cross-domain interactions more naturally.

The Head of AI at Synera articulated the vision: ``Future AI must reason about both individual components and larger systems with less data.'' This emphasis on data efficiency proves crucial given the scarcity of high-quality engineering training data. Rather than requiring millions of examples to learn basic physical principles, future systems might incorporate physics-informed architectures, symbolic reasoning capabilities, or hybrid approaches that combine neural networks with traditional simulation and analysis methods.

Progress in this area would enable transformative applications: AI systems that can truly understand CAD models and reason about design intent, agents that can automatically identify manufacturability issues by reasoning about tool access and material constraints, and systems that can optimize designs across multiple coupled physics domains simultaneously. Such capabilities would shift AI from being a peripheral assistant to being a core participant in engineering design processes.

\paragraph{4. Comprehensive Data Transformation and Knowledge Capture Pipelines}

Engineering knowledge exists in countless formats: CAD files, PDFs, images, spreadsheets, videos, tribal knowledge in experts' heads, and making this information accessible to AI systems requires sophisticated transformation pipelines. 

The breakthrough needed combines technical capabilities (robust document parsing, image understanding, knowledge extraction from unstructured sources, automatic structuring of informal information) with processes for systematically capturing knowledge before it is lost. Several interviewees mentioned the aging workforce and the imminent retirement of experts whose knowledge exists primarily in their heads. Creating tools and processes to capture and formalize this knowledge before it disappears represents an urgent need.

Such data transformation capabilities would also help address the data scarcity problem for training domain-specific AI models. If organizations could systematically transform their accumulated engineering artifacts into structured, machine-readable forms, they would create valuable training datasets even without generating entirely new design examples. 

An executive at an multinational semiconductor company mentioned ``Everybody wants to use AI, but how? You can’t just hand data off to a data scientist. The user has to work very closely with the data team to make it useful.'' 

\paragraph{5. Embedded Cultural Change Mechanisms and Training Frameworks}

Technical breakthroughs alone will not ensure successful adoption; organizations need proven frameworks to build AI literacy, manage change, and foster appropriate AI use cultures. Although this may seem less technically exciting than innovations in model architecture, multiple interviewees emphasized cultural factors as primary barriers to adoption.

The needed breakthroughs include: evidence-based training curricula that efficiently build AI literacy across engineering populations with diverse backgrounds; proven change management frameworks tailored to engineering organizations' specific characteristics; mechanisms for identifying and developing internal AI champions who can support peers; and methods for creating ``aha moments'' that help individuals develop appropriate trust calibration.

A NASA engineer's observation frames the long-term challenge: ``It is pretty clear to me this is going to work... but what roles do we want to have, and how do we bring everybody along?'' This question extends beyond individual organizations to society-level decisions about how we want human engineers and AI systems to collaborate, what skills future engineers need, and how we ensure that AI augments rather than simply displaces engineering expertise.

\paragraph{6. Governance Frameworks That Seamlessly Integrate with Engineering Review Processes}

Rather than creating entirely new governance structures for AI, organizations need frameworks that extend and adapt existing engineering review processes---PDR, CDR, TRR, change boards, safety reviews---to accommodate AI components. These frameworks must address traceability (ensuring that every AI decision is logged and attributable), auditability (allowing after-the-fact review of AI actions), rollback mechanisms (enabling reversal of AI changes that prove problematic), and graduated autonomy (defining which decisions AI can make independently versus which require human approval).

Such frameworks would ideally be developed collaboratively across industries rather than each organization creating incompatible internal approaches. Regulatory bodies, standards organizations, and industry consortia all have potential roles to play in developing and promulgating governance best practices that could accelerate adoption while maintaining appropriate safety and quality standards.

\section{How Agentic AI is Being Adopted: Present and Future} 
\label{sec: adoption-dynamics}

Our interviews revealed several adoption dynamics that explain both how AI systems are currently being integrated into engineering and manufacturing workflows and what conditions shape their continued adoption. Below, we outline the key themes, supported by insights from industry practitioners.

\begin{itemize}
    \item Adoption varies across and within organizations, and variety stems from several factors including understanding of AI and its capabilities, digital maturity, and organizational scale.
    \item Organizational buy-in has great impact on the use and effectiveness of AI systems, particularly buy-in at upper levels. 
    \item Preference for augmentation rather than replacement; humans are in the loop. 
    \item Provable reliability is necessary for full adoption. 
    \item Data security, privacy, and compliance requirements significantly constrain deployment options.
    \item Legacy toolchains and digital infrastructure readiness mediate the pace of adoption.
\end{itemize}

These themes are consistent with prior research showing that AI adoption is shaped less by model availability and more by organizational digital maturity, complementary infrastructure, workforce capabilities, and governance structures \citep{Sharma2020ReimaginingDiffusion,McElheran2024AIAdoptionAmerica}. In engineering and manufacturing contexts, recent work further emphasizes that higher-value applications emerge only when AI systems are integrated into existing toolchains, supported by structured data, and deployed within verification frameworks that enable trust in high-consequence environments \citep{ferdous2024automation}.

\subsection{Adoption Varies Across and Within Organizations}

AI adoption differed substantially not only between companies but also across teams, roles, and individuals within the same organization. Differences in digital maturity, familiarity with AI capabilities, workflow constraints, and personal attitudes toward automation all contribute to these variations. This heterogeneity mirrors large-scale evidence that AI adoption is highly uneven across firms and is strongly correlated with prior investments in cloud infrastructure, data assets, and innovation-oriented organizational strategies \citep{McElheran2024AIAdoptionAmerica}. Diffusion research similarly highlights that adoption occurs simultaneously at organizational and individual levels, where trust, skills, and perceived usefulness shape how technologies spread through local work practices \citep{Sharma2020ReimaginingDiffusion}.

In our interviews, we found that at the organizational level, companies ranged from highly mature adopters to those still in exploratory phases. A research group head at a global technology company emphasized that organizations often seek incremental improvements: ``Big businesses are looking, so far, only for optimization. Everybody is asking for improvement.'' They stressed that AI adoption follows an evolutionary sequence: ``First automation, then digitalization, and then AI. We cannot skip the evolution.'' Furthermore, while more and more people know how to use AI tools, very few are able to build it from scratch. As explored in Section~\ref{sec:ai-literacy}, an executive at a midsized engineering firm attributes the firm's competitive advantage to the ability to build such systems. 

Even within mature organizations, adoption is uneven. As described in Section~\ref{sec: building-trust-orgs} there are personas who sit at the extremes of AI adoption (enthusiasts and skeptics) and those in the middle ground (pragmatic users), and more effective AI implementations can happen from that middle ground. This typology appeared repeatedly across interviews. An engineer at an AI development company highlighted that there are ``tinkerers'' at large OEMs who experiment with AI independently, while others remain constrained by procedural or cultural inertia. Conversely, some engineers expressed deep skepticism. A back-end engineer at an engineering startup summarized this sentiment bluntly: ``I don’t use Copilot, I’d rather have a person design it because how is the AI supposed to know what the human experience actually is?''

Organizations responded to this heterogeneity by cultivating internal champions, communities of practice, and structured upskilling initiatives. The organizational culture spectrum is explored in Section~\ref{sec: org-culture-spectrum}, and building trust through organizational culture is explored in Section~\ref{sec: building-trust-orgs}. As discussed in both sections, our interviews suggest that organizational readiness sets the boundary conditions for adoption, but individual personas and local culture ultimately determine how AI diffuses through engineering workflows. The importance of internal champions and structured upskilling aligns with worker-centered studies showing that employees are often eager to engage with new technologies but require training and organizational support to translate experimentation into sustained productivity gains \citep{Armstrong2024AutomationWorkers}.

\subsection{Organizational Buy-In Strongly Influences Adoption}
\label{subsec:org_buy-in_adoption}

Leadership support consistently emerged as one of the most powerful determinants of AI adoption. Interviewees stressed that upper-level buy-in affects not only resource allocation but also the perception of AI’s legitimacy within engineering teams. Engineers reported that even technically strong tools fail to gain traction without clear executive endorsement, while enthusiastic leadership can rapidly accelerate interest and experimentation. In several cases, internal momentum was driven by influential early adopters or by leaders who actively championed AI-enabled workflows. Organizational buy-in therefore functions as a critical multiplier: without it, even compelling tools struggle to integrate; with it, adoption can propagate quickly.

This finding is in line with prior research on the adoption of AI tools in engineering and manufacturing firms. A 2023 study titled \textit{AI Adoption in America: Who, What, and Where} found that firms pursuing process innovation, holding patents, and targeting high growth are significantly more likely to adopt AI, highlighting the central role of managerial priorities in moving AI from experimentation into production use~\citep{McElheran2024AIAdoptionAmerica}. Furthermore, Alam et al. propose a development framework for the use of generative AI in design and manufacturing, but note that succeeding in that framework ``will require broad-based buy-in from business leaders,
operators, researchers, engineers, and policymakers''~\citep{ferdous2024automation}. 

\subsection{Preference for Augmentation Rather Than Replacement}

Across the vast majority of interviews, engineers expressed a clear preference for AI systems that augment rather than replace human expertise. In safety-critical and high-consequence engineering domains, human oversight is viewed as essential. A defense engineer summarized this succinctly: ``AI suggestions should be supplementary, not blindly trusted.'' An executive at AllSpice similarly emphasized that AI in PCB design serves as ``a co-pilot, not an autonomous decision-maker.''

This preference reflects both engineering culture and current technical limitations. Many teams intend to use AI for tasks such as preliminary design exploration, manufacturability checks, simulation setup, or documentation assistance, but want humans remain responsible for validating outputs and making final decisions. Synera's Head of AI noted that engineers increasingly seek creative collaboration with AI, ``Everybody wants to feel creativity with these models,'' but do not expect fully automated design anytime soon. Current patterns are already \textit{integrating} AI into human workflows. The next logical step that companies like Synera work towards is ``not about replacement, but rather turning AI into `coworkers' that work along side human engineers and perform entire workflows while handling the given challenges with human-like autonomy.''

These observations are consistent with prior research showing that AI adoption in complex, high-skill domains follows an augmentation trajectory in which systems first accelerate human workflows such as information synthesis, design iteration, and validation, while leaving responsibility for high-consequence decisions with domain experts \citep{ferdous2024automation}. Worker-centered studies further demonstrate that when new technologies are deployed in ways that enhance autonomy, learning opportunities, and safety, employees are significantly more likely to view them as complementary rather than substitutive \citep{Armstrong2024AutomationWorkers}. Moreover, involving workers directly in defining use cases and redesigning workflows has been shown to shift AI deployment toward augmentation and improve both organizational performance and job quality \citep{Kochan2024Bringing}.

\subsection{Prerequisites for Full Adoption}
\label{subsec:prereq_adoption}

\paragraph{Provable Reliability and Validation} In Section~\ref{sec: explainability}, we explored the need for explainability and verification in engineering workflows. As our interviews revealed, high-stakes engineering work requires AI systems to demonstrate reliability through rigorous validation. Across interviews, organizations emphasized that adoption depends not only on performance but on a system’s ability to produce \textit{verifiable} and \textit{auditable} outputs. A defense engineer explained: ``It would need to seem reasonable and testable, having a way to validate results against ground-truth cases would make me more comfortable using it.''

Enterprise organizations echoed these expectations. An engineer at a major defense contractor described efforts to impose the same checkpoints on AI systems that humans undergo during engineering reviews, while an engineer at Autodesk emphasized the need for reasonable, quantifiable, and repeatable accuracy. A research head at a global technology company echoed the importance of this, particularly with agentic systems where performance is not necessarily binary. In these cases we still need measurable performance checks at specified checkpoints, the same checkpoints that are likely already in place for human engineers. NASA engineers similarly noted that trust is built when AI tools perform reliably on established benchmark tasks, the same tasks used to onboard new engineers. Without this provable reliability, AI tools remain confined to low-stakes or advisory roles.

\paragraph{Data Security, Privacy, and Compliance Constrain Deployment}

As discussed in Section~\ref{sec: data-privacy}, data control emerged as one of the most significant constraints on AI adoption across engineering and manufacturing organizations. Many teams operate under strict regulatory regimes, such as ITAR, CMMC, or customer-imposed IP restrictions—that prohibit cloud-connected models or external data sharing. 
% One AI developer described this challenge succinctly: “No two similar companies want their data on the same server. BMW doesn’t want their parts on the same machine as Volkswagen.”
\begin{myquotebox}
No two similar companies want their data on the same server. BMW doesn’t want their parts on the same machine as Volkswagen.
\attributed{AI developer for engineering.}
\end{myquotebox}
Similarly, a leader at PTC highlights the risk of external data sharing to train LLM models ``Even between collaborating companies, there’s a lot of protection of data. They don’t want a first-mover disadvantage where they lose their IP.''

The director of structural dynamics and environments for a large aerospace company noted that ``all tools must be self-contained due to export controls and proprietary data concerns'' and an employee from a startup developing foundation AI models reported that many enterprise customers request fully local or on-device deployments: ``You don’t want your data stored in a cloud where somebody else can access it''. This trend extends to other domains including automated pharmaceutical production, with one company citing security as a reason for hesitation, noting that ``the stakes are high'' in their regulated field. These requirements significantly shape the technical architectures of AI tools, often necessitating on-premise, air-gapped, or edge-based deployments. In many cases, security and compliance, not performance, are the dominant barriers to adoption.

\paragraph{Legacy Toolchains and Digital Infrastructure Shape the Pace of Adoption}

Finally, as discussed in Section~\ref{sec: legacy-tools}, organizations repeatedly described integration challenges stemming from legacy software ecosystems. Many CAD, CAE, PLM, and documentation tools were not designed with modern APIs, modularity, or machine-interoperable interfaces. As an AI developer observed, “If you look at a CAE software that came in 1960, nobody thought we need a REST API endpoint so that in 2025 agents need this.” This creates a mismatch between the rapid evolution of AI systems and the slower update cycles of industrial software.

These constraints can limit the ability of agentic systems to orchestrate multi-step workflows or interact seamlessly with existing tools. However, adapting to them can also yield opportunities. AllSPice's CTO notes ``[something that has been] key to agentic design in hardware is embedding our AI tool in the engineering workflow in a way that makes it easy for engineers to quickly get full context and traceability. [This] is something you don’t fully get from a general-purpose tool, like ChatGPT, where you have to upload all of the data, explain all of the context, then export for any iteration.'' We observe that companies with modern, API-rich toolchains are able to adopt AI more quickly, while those with deeply entrenched legacy systems face heavier integration burdens. This dynamic highlights a key structural factor in AI adoption: the technical readiness of the software environment often determines the pace at which organizations can incorporate AI, independent of interest or capability.

\section{Limitations and Looking Forward}
This study is qualitative and intends to reflect the perspectives of the interviewed sample. Findings are intended to surface cross-cutting patterns, constraints, and design implications rather than to estimate prevalence, benchmark tools, or quantify causal impacts of AI adoption. Due to limited sample size and non-randomized sampling process, some of the observations may be limited in scope and should be interpreted carefully. In addition, interview data are shaped by participants’ roles, experiences, and exposure to AI tools, which may bias which challenges or opportunities they emphasize. The rapid pace of AI development also means that some observations may evolve as tools, governance frameworks, and industry practices mature. As such, these findings should be interpreted as a state-of-practice snapshot intended to inform future inquiry rather than a comprehensive or generalizable assessment.

\section{Conclusion}

This study set out to characterize the current state and near-term trajectory of AI adoption in engineering and manufacturing workflows. Three cross-cutting findings emerged from 33 interviews across 28 organizations.
First, AI utility in engineering today clusters around structured, repetitive, and data-intensive tasks. These are domains with clear success criteria, established patterns, and manageable consequences of failure. Higher-value agentic gains—orchestrating multi-step workflows, enabling rapid design exploration, supporting real-time decision-making in complex systems—are emerging but require infrastructure, verification frameworks, and data quality that are largely absent today.

Second, the primary constraints on adoption are not model capability but ecosystem readiness: fragmented, machine-unfriendly data; legacy toolchains without programmatic interfaces; stringent security and regulatory requirements; and the absence of engineering-grade verification standards. Each of these is addressable, but none has an easy or immediate solution. Progress requires coordinated investment across AI developers, tool vendors, engineering firms, and standards bodies.
Third, governance is not a temporary barrier to be overcome en route to automation—it is a permanent feature of responsible AI deployment in high-stakes environments. The organizations achieving the most effective AI adoption share several characteristics: they maintain human-in-the-loop oversight for safety-critical applications, architect systems to meet security requirements, prioritize explainability and determinism, align AI governance with existing engineering review processes, and invest in cultural trust-building through communities of practice and peer learning.
For engineering organizations, these findings suggest prioritizing data infrastructure, verification practices, and workforce AI literacy over one-off tool deployments. The most durable gains will come from treating AI adoption as an organizational transformation, not a software procurement decision.

For AI and CAD/CAM/CAE vendors, the most impactful investments lie in open APIs, standardized agent interfaces, on-premise deployment options, and tools that expose their reasoning and data sources in ways that build rather than erode engineering trust.
For the research community, the findings point to several open problems: verification frameworks for probabilistic systems operating in safety-critical engineering contexts; physics-informed architectures with genuine multi-domain reasoning capabilities; data transformation pipelines capable of extracting structured knowledge from the heterogeneous artifacts of engineering practice; and governance models that scale with the speed of agentic AI without sacrificing auditability.

This study is qualitative and reflects the perspectives of the interviewed sample. Findings are intended to surface cross-cutting patterns, constraints, and design implications rather than to estimate prevalence, benchmark tools, or quantify causal impacts. The rapid pace of AI development means some observations will evolve as tools, governance frameworks, and industry practices mature. These findings should be interpreted as a state-of-practice snapshot intended to inform future inquiry rather than a comprehensive or generalizable assessment.

\section*{Acknowledgments}
The authors thank the MIT Initiative for New Manufacturing for its support and guidance of this work. We extend our sincere gratitude to Suzanne Berger, Ben Armstrong, Julie Diop, and David Mindell for their expertise, insights, and feedback during the early stages of this project. Finally, we thank the interview participants whose insights and experiences form the backbone of this work.

\bibliographystyle{plainnat} %plainnat} %unsrtnat
\bibliography{main}

\newpage
\appendix
\section{Interview Questions}
\label{sec: interview_questions}

\begin{longtable}{p{0.28\textwidth} p{0.68\textwidth}}
\caption{The interview guide that our team developed and worked from.} \\
\toprule
\textbf{Interviewee Type} & \textbf{Question} \\
\midrule
\endfirsthead

\toprule
\textbf{Interviewee Type} & \textbf{Question} \\
\midrule
\endhead

\midrule
\multicolumn{2}{r}{\textit{Continued on next page}} \\
\midrule
\endfoot

\bottomrule
\endlastfoot

\multicolumn{2}{l}{\textbf{For All Interviews}} \\
& What does your company do? \\
& Who are your typical customers? \\
& What is your role? \\
& What is your day-to-day work like? \\
& Have you heard of Generative AI (e.g., ChatGPT, Copilot) applied to engineering tasks? \\
& How familiar are you with the concept of AI agents or agentic AI? \\
& Have you experimented with any form of agent-based systems? \\

\addlinespace
\multicolumn{2}{l}{\textbf{AI Agent Creators}} \\
& What’s the origin story of your agent-based system? What problem were you trying to solve? \\
& What types of environments (e.g., CAD, PLM, ERP) does your agent operate in? \\
& What sets an “agent” apart from a copilot or script in your architecture? \\
& Which engineering use cases are most tractable right now? Which are furthest away? \\
& What data sources are required to train or fine-tune your agents effectively? \\
& Do you anticipate your system being embedded in third-party tools or acting as a layer on top? \\
& What breakthroughs are needed for multi-agent collaboration in engineering contexts? \\
& How is tool usage changing across user demographics? \\
& What characteristics of CAD/CAM/CAE tools make them compatible with agents? \\
& What are the key data privacy and security concerns? \\

\addlinespace
\multicolumn{2}{l}{\textbf{CAD/CAM/CAE Providers}} \\
& How do you currently integrate your software into users’ larger workflows? \\
& How do you enable interoperability between your tools and external systems? \\
& Are you exploring AI agent-based extensions for your software suite? \\
& How do you envision agentic AI fitting into your product roadmap? \\
& Where would AI agents create the most value: design, simulation, or post-processing? \\
& What role do APIs or SDKs play in advanced workflows? \\
& Are you considering embedding language models or agents natively? \\
& What are the biggest technical or business risks in deploying AI agents inside commercial tools? \\

\addlinespace
\multicolumn{2}{l}{\textbf{Engineering or Manufacturing Enterprises}} \\
& What is your team’s level of autonomy vs. use of external firms? \\
& Which engineering tasks require the most trial and error? \\
& What knowledge currently “lives in people’s heads” vs. being codified? \\
& Do engineers use AI tools like ChatGPT or Copilot in daily work? How? \\
& Walk me through your typical workflow - are there any repetitive tasks? \\
& If an AI agent could handle one task, what would it be? \\
& How is your engineering data stored today (cloud, local servers, individual machines)? \\
& Would AI-generated suggestions be trusted? Why or why not? \\
& How do engineering tools and workflows vary across business units? \\
& Are there existing automation efforts (MBSE, digital twins) agents could build on? \\
& Are pilot programs for agent-based systems underway? \\
& What is the appetite for autonomous agents in safety-critical workflows? \\
& How modular or siloed are engineering and production data systems? \\
& What security, compliance, or regulatory requirements apply to AI decision support? \\
& What breakthroughs are needed for multi-agent collaboration in engineering contexts? What are you excited about? What's next? \\
& Are there fully automated tasks today? \\
& Where do you see AI agents creating the most value for your customers? \\
% &       Simulation automation (mesh, boundary condition setup)? \\
% &       Parametric design exploration? \\
% &       Toolpath generation and optimization? \\
% &       Robotic assembly coordination? \\
% &       Predictive maintenance? \\
% &       Text generation from legacy documentation? \\
% &       Defect detection in large-scale additive manufacturing? \\
& \hspace{1em} Simulation automation (mesh, boundary condition setup)? \\
& \hspace{1em} Parametric design exploration? \\
& \hspace{1em} Toolpath generation and optimization? \\
& \hspace{1em} Robotic assembly coordination? \\
& \hspace{1em} Predictive maintenance? \\
& \hspace{1em} Text generation from legacy documentation? \\
& \hspace{1em} Defect detection in large-scale additive manufacturing? \\
& Which of these do you see as most urgent for your business? Why? \\
& How do you go about identifying areas where software could improve user experience or augment their performance? \\

\addlinespace
\multicolumn{2}{l}{\textbf{For All: Looking Forward}} \\
& How do you expect work to be affected by GenAI or agentic AI technologies in the next 1–3 years? \\
& What would success look like if AI agents were integrated into your engineering workflow? \\

\end{longtable}

\end{document}